\documentclass[10pt,journal]{IEEEtran}

\ifCLASSOPTIONcompsoc
  \usepackage[nocompress]{cite}
\else
  \usepackage{cite}
\fi

\ifCLASSINFOpdf
\else
\fi

\usepackage[utf8]{inputenc} 
\usepackage[T1]{fontenc}    
\usepackage{hyperref}       
\usepackage{url}            
\usepackage{booktabs}       
\usepackage{amsmath}
\usepackage{amssymb}
\usepackage{amsfonts}       
\usepackage{nicefrac}       
\usepackage{microtype}      
\usepackage{xcolor}         
\usepackage{enumitem}
\usepackage{musicography}
\usepackage{subcaption}
\usepackage{bm}
\usepackage{multicol}
\usepackage{multirow}
\usepackage{graphicx}
\usepackage{comment}

\usepackage{array}
\makeatletter

\newcommand*{\@rowstyle}{}

\newcommand*{\rowstyle}[1]{
  \gdef\@rowstyle{#1}%
  \@rowstyle\ignorespaces%
}

\newcolumntype{=}{
  >{\gdef\@rowstyle{}}%
}

\newcolumntype{+}{
  >{\@rowstyle}%
}

\newcommand{\revision}[1]{{\color{black}#1}}
\newcommand{\revrdtwo}[1]{{\color{black}#1}}
\newcommand{\revtbrow}{\rowstyle{\color{black}}}

\begin{document}

\title{MuseMorphose: Full-Song and Fine-Grained \revrdtwo{Piano} Music Style Transfer with One Transformer VAE}

\author{Shih-Lun~Wu, 
        and~Yi-Hsuan~Yang,~\IEEEmembership{Senior~Member,~IEEE}
\IEEEcompsocitemizethanks{\IEEEcompsocthanksitem Both SL and YH were with the Taiwan AI Labs. 
Besides, SL was also  with the Department of CSIE, National Taiwan University; and YH was also  with the Research Center for IT Innovation, Academia Sinica. 
\protect\\
E-mail: b06902080@csie.ntu.edu.tw; yhyang@ailabs.tw
}
\thanks{Manuscript received Oct'21; revised Jul'22 \& Oct'22; accepted Dec'22.}
}

\markboth{IEEE/ACM Transactions on Audio, Speech, and Language Processing,~Vol.~\#, No.~\#, }%
{Shell \MakeLowercase{\textit{et al.}}: Bare Demo of IEEEtran.cls for Computer Society Journals}



\IEEEtitleabstractindextext{%
\begin{abstract}
Transformers and variational autoencoders (VAE) have been extensively employed for symbolic (e.g., MIDI) domain music generation.
While the former boast an impressive capability in modeling long sequences, the latter allow users to willingly exert control over different parts (e.g., bars) of the music to be generated.
In this paper, we are interested in bringing the two together to construct a single model that exhibits both strengths.
The task is split into two steps.
First, we equip Transformer decoders with the ability to accept segment-level, time-varying conditions during sequence generation.
Subsequently, we combine the developed and tested in-attention decoder with a Transformer encoder, and train the resulting MuseMorphose model with the VAE objective to achieve style transfer of long 
\revision{pop piano pieces},
in which users can specify musical attributes including rhythmic intensity and polyphony (i.e., harmonic fullness) they desire, down to the bar level.
Experiments show that MuseMorphose outperforms recurrent neural network (RNN) based baselines on numerous widely-used metrics for style transfer tasks.
\end{abstract}

}

\maketitle

\IEEEdisplaynontitleabstractindextext

%
\IEEEpeerreviewmaketitle

\section{Introduction}\label{sec:introduction}

Automatic music composition, i.e., the generation of musical content in symbolic formats such as MIDI,\footnote{Check \url{https://midi.org/specifications} for detailed specifications.} has been an active 
research topic with endeavors dating back to more than half a century ago \cite{hiller1959experiment}.
Due to the renaissance of neural networks, 
we have seen in recent years a proliferation of deep learning-based methods for music composition \cite{hadjeres2017deepbach, dong2018musegan,ji20survey}. 
\revision{For such tasks, a commonly followed pipeline is to first transcribe recorded music performances \cite{hawthorne2018onsets} into sheet music, which can be stored as MIDI files, and then translate MIDI events into sequences of tokens \cite{oore2018time, huang2020pop} that can be fed into neural sequence models.
The latter step resembles how researchers in natural language processing (NLP) represent sentences of words with encodings \cite{sennrich2016neural}.}
Different network architectures have been employed depending on the target applications, 
but the state-of-the-art models are usually based on one of the following two architectures
\cite{huang19music, donahue2019lakhnes, roberts2018hierarchical}---Transformers \cite{vaswani2017attention} and variational autoencoders (VAE) \cite{kingma2014auto}.

\begin{figure}
    \centering
    \includegraphics[width=\columnwidth]{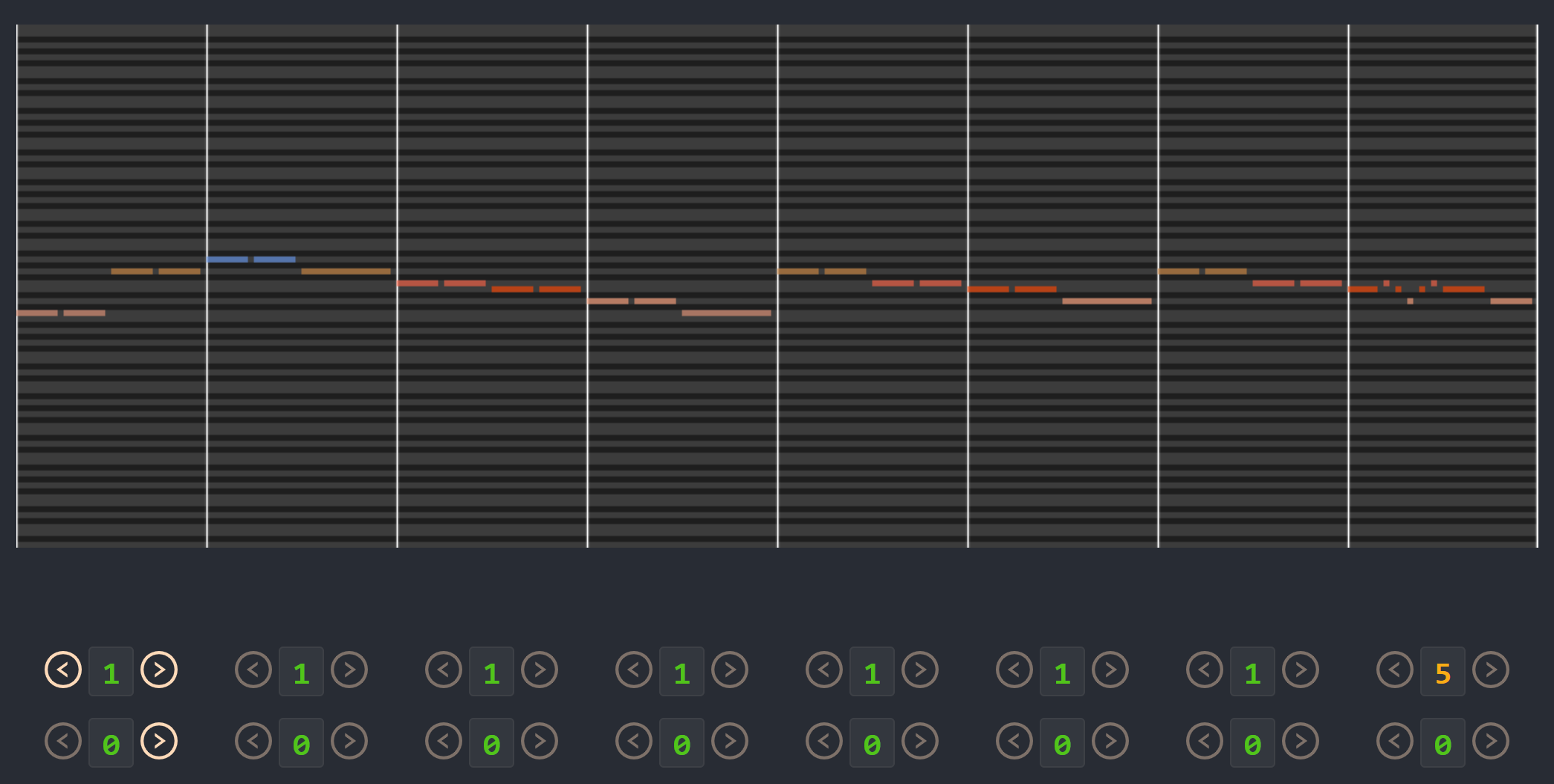} \\
    \vspace{1mm}
    \includegraphics[width=\columnwidth]{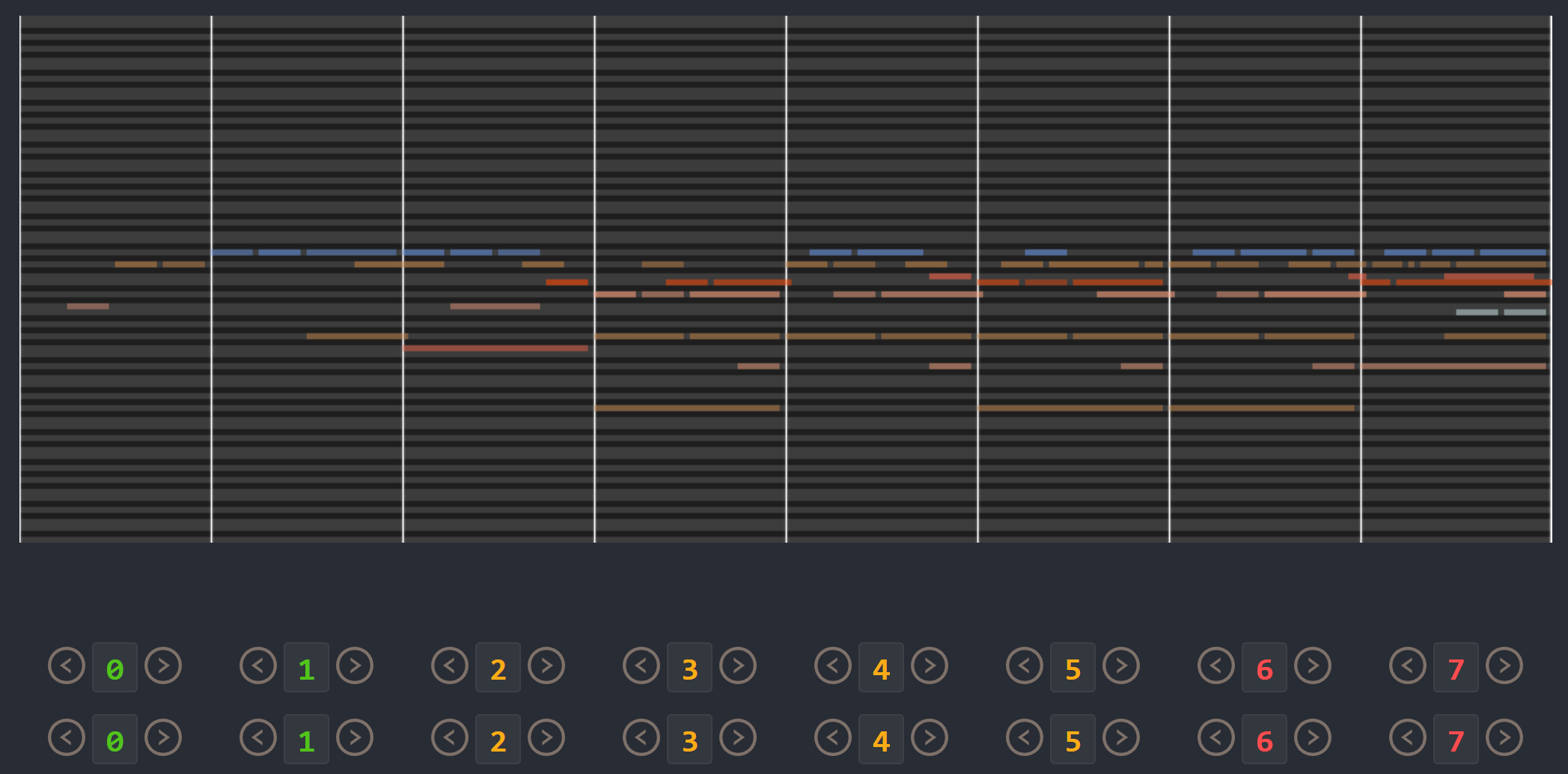}
    \caption{Visualizations of an original music excerpt (top) and one of its style-transferred generations by \textit{MuseMorphose} (bottom). Digits in the upper and lower rows represent our target attributes being controlled, namely, bar-wise \textit{rhythmic intensity} and \textit{polyphony} classes, respectively.}
    \label{fig:demo}
\end{figure}

Transformers \cite{vaswani2017attention} are a family of neural sequence models that are 
generally 
thought of as the potent successors of recurrent neural networks (RNN) \cite{hochreiter97lstm, cho-etal-2014-properties}.
Thanks to 
their self-attention mechanism
to aggregate information from 
the hidden states of all previous tokens in a sequence,
Transformers are able to 
compose coherent music of \revrdtwo{3$\sim$5} minutes long \revrdtwo{(which mostly contains over 2K sequence tokens)}, as demonstrated by \textit{Music Transformer} \cite{huang19music} and \textit{MuseNet} \cite{payne2019musenet}.
Variants of Transformers 
have been employed to model multi-track music \cite{donahue2019lakhnes}, pop piano performances \cite{huang2020pop}, jazz lead sheets \cite{wu2020jazz}, and even guitar tabs \cite{chen2020automatic}. 
Objective and subjective evaluations in \cite{huang19music, donahue2019lakhnes} and \cite{dai2019transformer} have all proven Transformers to be the state-of-the-art model for long sequence generation.

VAEs \cite{kingma2014auto} are a type of deep latent variable model comprising an encoder, a decoder, and a Kullback-Leibler (KL) divergence-regularized latent space in between.
\revision{While neural sequence encoders/decoders, like RNNs and Transformers, gather and consume information from tokens in a sequence, VAE's training objective restricts the amount of information the latent space may store. This induces better structuring of information in the latent space, and also leaves the sequence decoder with enough uncertainty to be creative.}
The \revision{resulting} strength of VAEs is that they grant human users access to the generative process through operations on the learned latent space \cite{roberts2018hierarchical}, 
where
compact semantics of a musical excerpt are stored,
making them a great choice for tasks like controllable music generation and music style (or, attribute) transfer.
For example, \textit{MusicVAE} \cite{roberts2018hierarchical} and \textit{Music FaderNets} \cite{tan2020music} showed respectively that low-level (e.g., note density) and high-level (e.g., arousal level) musical features of machine compositions can be altered via latent vector arithmetic.
\textit{Music FaderNets}, \textit{GLSR-VAE} \cite{hadjeres2017glsr}, and \textit{MIDI-VAE} \cite{brunner2018midi} imposed auxiliary losses on the latent space to force some latent dimensions to be discriminative of the musical style or attributes to be controlled.
In contrast, Kawai \textit{et al.} \cite{kawai2020attributes} used adversarial learning \cite{goodfellow2014generative} to prevent the latent space from encoding attribute information, and delegated attribute control to learned embeddings fed to the decoder.


The aforementioned VAE models are, however, all based on RNNs, 
whose capabilities in modeling long sequences are known to be limited.
On the other hand,
conditional generation with Transformers are usually done by providing extra tokens.
For instance, MuseNet \cite{payne2019musenet} used composer and instrumentation tokens, prepended to the event sequence, to affect the composition's style and restrict the instruments used. Sometimes, Transformers are given a composed melody (also as a token sequence) and are asked to generate an accompaniment.
Such tasks have been addressed with encoder-decoder architectures \cite{choi2020encoding, ren2020popmag}, or by training a decoder with interleaved subsequences of the melody and the full piece \cite{hsiao21aaai}.
The two scenarios above can be seen as two extremes in terms of the restrictiveness of given conditions.
Users may either only be able to decide the broad categories (e.g., genre, composer, etc.) of the generation,
or have to be capable of coming up with a melody themselves.
The middle ground of constraining the generation with high-level flow of musical ideas in the form of latent vectors, with which users may engage easily and extensively in the machine creative process as achieved by RNN-based VAEs, has not yet been studied for Transformers to the best of our knowledge.



The goal of this paper is therefore to construct such a model that attains all the aforementioned strengths of Transformers and VAEs.
We take two steps to achieve the goal.
First, we devise mechanisms to condition Transformer decoders \cite{radford2019language} with segment-level (i.e., \textit{bar}-level in our case), time-varying conditioning vectors during long sequence generation.
Three methods, i.e., \textit{pre-attention}, \textit{in-attention}, and \textit{post-attention}, which inject conditions into Transformers 
\revision{\textit{before}, \textit{throughout}, or \textit{after} the attention layers respectively,}
are proposed
\revision{(implementation details in Sec.~\ref{sec:transformers})}.
We conduct an objective study to show that in-attention most effectively exerts control over the model.
Next, we combine an in-attention-equipped Transformer decoder with a Transformer encoder \cite{devlin18bert}, 
which learns to extract each bar's high-level blueprint independently,
to build our ultimate \textit{MuseMorphose} model for fine-grained music style transfer. This encoder-decoder network is trained with the VAE objective \cite{kingma2014auto}, and enables music style transfer through attribute embeddings \cite{fu2018style}.
Experiments demonstrate that MuseMorphose excels in generating style-transferred versions of pop piano performances (i.e., music with expressive timing and dynamics, instead of plain sheet music) of 32 bars (or measures) long, in which users can freely control two ordinal musical attributes of each bar: \textit{rhythmic intensity} and \textit{polyphony}, \revision{which roughly represent the density of notes seen from the time/pitch axis respectively (see Eqs.~(\ref{eq:rhym-int}) and (\ref{eq:polyphony}) for formal definitions)}.

\revision{These two attributes are chosen for they largely affect the music's feeling (e.g., agitated, calm, etc.) and are hence easily perceptible \revrdtwo{\cite{panda2018novel, panda2020audio}}.
Besides, we set the granularity of conditions to a bar for several reasons.
From a musical perspective, bars are the basic unit of recurring strong/weak beat cycles, serving as the basis of all rhythmic patterns, but they are also long enough to expose a musical idea.
Sudden changes of emotion often happens on boundaries between bars, too.
From a machine learning viewpoint, we find that
the \textit{number of bars in a piece}, and the \textit{number of tokens in a bar} share a similar order of magnitude (averaging at about 100 and 50, respectively, on our pop piano dataset),
thus making bars a suitable target for learning intermediate representations.}

Parameterizing VAEs with Transformers is a less explored direction. 
\textit{Optimus} \cite{li2020optimus} successfully linked together a pair of pre-trained Transformer encoder (\textit{BERT}) \cite{devlin18bert} and decoder (\textit{GPT-2}) \cite{radford2019language} with a VAE latent space, and devised mechanisms for the latent variable to exert control over the Transformer decoder.
Linking Transformers and VAE for story completion \cite{wang2019t} has also been attempted.
However, both works above were limited to generating short texts (only one sentence), and conditioned the Transformer decoder at the global level.
Our work extends the approaches above to tackle the case of long sequences, maintains VAEs' fine-grained controllability, and verifies that such control can be achieved without a discriminator and auxiliary losses.

Our key contributions can be summarized as:
\begin{itemize}[leftmargin=*]
    \item We devise the \textit{in-attention} method to firmly harness Transformers' generative process with segment-level conditions.
    \item We pair a Transformer decoder featuring in-attention with a bar-level Transformer encoder to form MuseMorphose, our Transformer-based VAE, which marks an advancement over \textit{Optimus} \cite{li2020optimus} in terms of conditioning mechanism, granularity of control, and accepted sequence length.
    \item We employ MuseMorphose for the style transfer of 32-bar-long\footnote{We note that MuseMorphose can actually take arbitrarily long inputs.} pop piano performances, on which it surpasses
    state-of-the-art RNN-based VAEs
    \cite{brunner2018midi, kawai2020attributes} on various metrics without having to use style-directed loss functions.
\end{itemize}

Fig.~\ref{fig:demo} displays one of MuseMorphose's compositions,\footnote{Each tone in the chromatic scale (C, C\#, \dots, B) is colored differently.} in which the model responds precisely to the increasing rhythm intensity and polyphony settings, while keeping the contour of the original excerpt well.
We encourage readers to visit our companion website\footnote{Companion website: \url{slseanwu.github.io/site-musemorphose}} to listen to more generations by MuseMorphose.
Moreover, we have also open-sourced our implementation of MuseMorphose.\footnote{Open-source code: \url{github.com/YatingMusic/MuseMorphose}}
\revrdtwo{In the interest of space, this paper contains some pointers to extra tables and figures released as online supplemental materials.}\footnote{Supplemental: \url{slseanwu.github.io/site-musemorphose/assets/supplement.pdf}}

The remainder of this paper is structured as follows.
Section 2 provides a comprehensive walk-through of the technical background.
Sections 3 and 4, which are the main body of our work, focus respectively on the segment-level conditioning for Transformer decoders, and the MuseMorphose Transformer-based VAE model for fine-grained music style transfer.
In each of these two sections, we start from problem formulation; then, we elaborate on our method(s), followed by evaluation procedure and metrics; finally, we present the results and offer some discussion.
Section 5 concludes the paper and provides potential directions to pursue in the future.

\section{Technical Background}
\subsection{Event-based Representation for Music}\label{sec:ev-repr}
To facilitate the modeling of music with neural sequence models, an important pre-processing step is to tokenize a musical piece $X$ into a sequence of events, i.e., $X = \{x_1, x_2,\dots,x_T\}$,
where $T$ is the length of the music in the resulting event-based representation. 
Such tokenization process can be done straightforward with music in the symbolic format of MIDI.
A MIDI file stores a musical piece's tempo (in beats per minute, or bpm), time signature (e.g., 3/4, 4/4, 6/8, etc.), as well as each note's onset and release timestamps and velocity (i.e., loudness).
Multiple tracks can also be included for pieces with more than one playing instruments.


There exist several ways to represent symbolic music as token sequences \cite{oore2018time, roberts2018hierarchical}.
The representations adopted in our work are based on Revamped MIDI-derived events (REMI) \cite{huang2020pop}.
It incorporated the intrinsic units of time in music, i.e., \textit{bars} and \textit{beats},
the former of which naturally define the segments to impose conditions. 
In REMI, a musical piece is represented as a sequence with the following types of tokens:
\begin{itemize}[leftmargin=*]
    \item \textsc{Bar} and \textsc{Sub-Beat} ($1{\sim}16$, denoting the time within a bar, in quarter beat, i.e., 16$^{\text{th}}$ note, increments) events denote the progression of time. 
    \item \textsc{Tempo} events explicitly set the pace (in beats per minute, or bpm) the music should be played at.
    \item \textsc{Pitch}, \textsc{Velocity}, and \textsc{Duration} events, which always co-occur at each note's onset,
    marks the pitch, loudness (in 32 levels), and duration (in 16$^{\text{th}}$ note increments) of a note.
\end{itemize}
The authors of REMI also suggested using rule-based chord recognition algorithms to add \textsc{Chord} events, e.g., \textsc{Cmaj}, \textsc{G\footnotesize{7}}, to inform the sequence model of local harmonic characteristics of the music.
To suit our use cases, we introduce some modifications to REMI, as explained in Sections 3 and 4.

\subsection{Transformers}

In general, a Transformer operates on an input sequence $X = \{x_1,\dots,x_T\}$.\footnote{Combinations of Transformers operating on multiple sequences are also often seen.}
A token $x_t$ in the input sequence first 
becomes a token embedding $\bm{x}_t \in \mathbb{R}^{d}$, where $d$ is the model's hidden state dimension. 
Then, $\bm{x}_t$ goes through the core of the Transformer---a series of $L$ \textit{(self-)attention modules} sharing the same architecture.
An attention module\footnote{It is also often called a \textit{(self-)attention layer}. However, we use the term \textit{module} here for it actually involves multiple layers of operations.} can be further divided into 
a \textit{multi-head attention} (MHA) \cite{vaswani2017attention} sub-module, and a \textit{position-wise feedforward net} (FFN) sub-module.
Suppose 
$\bm{x}_t$ has passed $l-1$ such attention modules to produce the hidden state $\bm{h}_t^{l-1} \in \mathbb{R}^{d}$, the operations that take place in the next (i.e., $l^{\text{\,th}}$) attention module can be summarized as follows:
\begin{align}
    \bm{s}_t^l &= \text{LayerNorm}(\bm{h}_t^{l-1} + \text{MHA}(\bm{h}_t^{l-1} \mid H^{l-1})) \\
    \bm{h}_t^l &= \text{LayerNorm}(\bm{s}_t^l + \text{FFN}(\bm{s}_t^l)) \label{eqn:tf-ffn},
\end{align}
where $H^{l-1} = [\bm{h}_1^{l-1},\dots,\bm{h}_T^{l-1}]^\top \in \mathbb{R}^{T \times d}$, and $\bm{s}_t^{l}, \bm{h}_t^{l} \in \mathbb{R}^{d}$. The presence of residual connections \cite{he2016deep} and layer normalization (LayerNorm) \cite{ba2016layer} ensures that gradients back-propagate smoothly into the network.
The MHA sub-module is the key component for each timestep to gather information from the entire sequence.
In MHA, first, the \textit{query}, \textit{key}, and \textit{value} matrices ($Q^l, K^l, V^l \in \mathbb{R}^{T \times d}$) are produced:
\begin{align}
    Q^l &= H^{l-1}W_Q^l &
    K^l &= H^{l-1}W_K^l &
    V^l &= H^{l-1}W_V^l,
\end{align}
where $W_Q^l, W_K^l, W_V^l \in \mathbb{R}^{d \times d}$ are learnable weights. 
$Q^l$ is further split into $M$ heads $[Q^{l, 1},\dots,Q^{l, M}]$, where $Q^{l, m} \in \mathbb{R}^{T \times \frac{d}{M}}, \forall m \in \{1,\dots,M\}$, and so are $K^l$ and $V^l$.
Then, the \textit{dot-product attention} is performed in a per-head fashion:
\begin{equation}
    \mathrm{Att}^{l, m} = \text{softmax}(\frac{Q^{l, m}{K^{l, m}}^{\top}}{\sqrt{d/M}})V^{l, m}
\end{equation}
where $\mathrm{Att}^{l, m} \in \mathbb{R}^{T \times \frac{d}{M}}$.
Causal (lower-triangular) masking is applied before softmax in autoregressive sequence modeling tasks.
The attention outputs (i.e., $\mathrm{Att}^{l, 1}, \dots, \mathrm{Att}^{l, M}$) are then concatenated and linearly transformed by learnable weights $W_O^l \in \mathbb{R}^{d \times d}$, thereby concluding the the MHA sub-module.

However, the mechanism above is permutation-invariant, 
which is clearly not a desirable property for sequence models.
This issue can be addressed by adding a \textit{positional encoding} to the input token embeddings, i.e.,
   $\bm{x}_t = \bm{x}_t + \mathrm{pos}(t)$, 
where $\mathrm{pos}(t)$ can either be fixed sinusoidal waves \cite{vaswani2017attention}, or learnable embeddings \cite{devlin18bert}. 
More sophisticated methods which inform the model of the relative positions between all pairs of tokens 
have also been invented \cite{shaw2018self, ke2021rethinking, wang2021position, liutkus2021relative}.


Due to attention's quadratic memory complexity
it is often impossible for Transformers to take sequences with over 2$\sim$3k tokens, leading to the context fragmentation issue.
To resolve this, \textit{Transformer-XL} \cite{dai2019transformer} leveraged
segment-level recurrence and hidden states cache to allow tokens in the current segment to refer to those in the preceding segment.
Algorithms to estimate attention under linear complexity have also been proposed recently \cite{katharopoulos2020transformers, choromanski2021rethinking}. 


\subsection{Variational Autoencoders (VAE)}\label{sec:vae-trick}


VAEs \cite{kingma2014auto}
are based on the hypothesis that generative processes follow $p(\bm{x}) = p(\bm{x} \vert \bm{z})p(\bm{z})$, where $p(\bm{z})$ is the prior of some intermediate-level latent variable $\bm{z}$.
It has been  shown that generative modeling can be achieved through maximizing:
\begin{equation}
    \mathcal{L}_\text{ELBO} = 
    \mathbb{E}_{\bm{z}\sim q_\phi(\bm{z}\vert\bm{x})}\log p_\theta(\bm{x}\vert\bm{z}) - D_\text{KL}(q_\phi(\bm{z}\vert\bm{x}) \| p(\bm{z})),\label{eqn:elbo}
\end{equation}
where $D_\text{KL}$ denotes KL divergence, and $q_\phi(\bm{z}\vert\bm{x})$ is the estimated posterior distribution emitted by the encoder (parameterized by $\phi$), while $p_\theta(\bm{x}\vert\bm{z})$ is the likelihood modeled by the decoder (parameterized by $\theta$).
It is often called the \textit{evidence lower bound} (ELBO) objective since it can be proved that:     $-\mathcal{L}_\text{ELBO} \leq \log p(\bm{x}).$
For simplicity, we typically restrict $q_\phi(\bm{z}\vert\bm{x})$ to isotropic gaussians $\mathcal{N}(\bm{\mu}, \text{diag}(\bm{\sigma}^2))$, and have the encoder emit the mean and standard deviation of $\bm{z}$, i.e., $\bm{\mu}, \bm{\sigma} \in \mathbb{R}^{\text{dim}(\bm{z})}$.
The prior distribution $p(\bm{z})$ is often set to be isotropic standard gaussian $\mathcal{N}(\bm{0}, I)$.

For practical considerations, $\beta$-VAE \cite{higgins2017beta} 
proposed that we may adjust the weight on the KL term of the VAE objective with an additional hyperparameter $\beta$.
Though using $\beta \neq 1$ deviates from optimizing ELBO, 
it grants practitioners the freedom to trade off reconstruction accuracy for a smoother latent space and better feature disentanglement (by opting for $\beta > 1$).
It is worth mentioning that, in musical applications, $\beta$ is often set to $0.1{\sim}0.2$ \cite{roberts2018hierarchical, brunner2018midi, kawai2020attributes}.

In the literature, it has been repeatedly mentioned that VAEs suffer from the \textit{posterior collapse} problem \cite{sonderby2016ladder, bowman2016generating, dieng2019avoiding}, in which case the estimated posterior $q_\phi(\bm{z}\vert\bm{x})$ fully collapses onto the prior $p(\bm{z})$, leading to 
an information-less latent space.
The problem is especially severe when the decoder is powerful, or when the decoding is autoregressive, where previous tokens in the sequence reveal strong information.
Numerous methods have been proposed to tackle this problem, either by tweaking the training objective \cite{kingma2016improving, fu2019cyclical} or slightly modifying the model architecture \cite{dieng2019avoiding}. 
Kingma \textit{et al.} \cite{kingma2016improving} introduced a hyperparameter $\lambda$ to the VAE objective
to ensure that each latent dimension may store $\lambda$ nats (1 nat $\approx$ 1.44 bits) of information without being penalized by the KL term.
\textit{Cyclical KL annealing} \cite{fu2019cyclical} periodically adjusts $\beta$ (i.e., weight on the KL term)
during training.
Empirical evidence have shown that this simple technique 
benefited downstream tasks such as language modeling and text classification.
\textit{Skip-VAE} \cite{dieng2019avoiding}, on the other hand, addressed posterior collapse with slight architectural changes to the model.
In Skip-VAEs, the latent condition $\bm{z}$ is fed to all layers and timesteps instead of just the initial ones.
Theoretically and empirically, it was shown that Skip-VAEs increases the mutual information between input $\bm{x}$ and the estimated latent condition $\bm{z}$.
In our work, we take advantage of all of the techniques above to facilitate the training of our Transformer-based VAE.

\section{Conditioning Transformer Decoders at the Segment Level}
\label{sec:part1}

This section sets aside latent variable models first, and 
focuses on
conditioning an autoregressive Transformer decoder with a series of pre-given, time-varying conditions $c_1,\dots,c_K$, which are continuous vectors each belonging to a pre-defined, non-overlapping segment of the target sequence.
We propose three mechanisms, namely, \textit{pre-attention}, \textit{in-attention}, and \textit{post-attention} conditioning, to approach this problem. 
Experiments demonstrate that, in terms of offering tight control, in-attention surpasses the other two, as well as a baseline mechanism that is slightly tweaked from \textit{Optimus} \cite{li2020optimus}.

\subsection{Problem Formulation}\label{sec:form-ch3}
Under typical settings, 
Transformer decoders \cite{vaswani2017attention} are employed for autoregressive generative modeling of sequences:
\begin{equation}
    p(x_t \vert x_{<t}) \,,
    \label{eq:uncond}
\end{equation}
where $x_t$ is the element of a sequence to predict at timestep $t$, and $x_{<t}$ is all the (given) preceding elements of the sequence. 
Once a model is trained, the model can generate new sequences autoregressively, i.e., one new element at a time based on all previously generated elements \cite{radford2019language}.

The \textit{unconditional} type of generation associated with Eq.~(\ref{eq:uncond}), however, does not offer control mechanisms to guide the generation process of the Transformer. 
One alternative is to consider a \textit{conditional} scenario where the model is informed of a \textit{global} condition $c$ for the entire sequence to generate, as used by, for example, the \textit{CTRL} model for text \cite{keskar2019ctrl} or the \textit{MuseNet} model for music \cite{payne2019musenet}: 
\begin{equation}
    p(x_t | x_{<t}, c) \,.
    \label{eq:cond_global}
\end{equation}
When the target sequence length is long, it may be beneficial to extend Eq.~(\ref{eq:cond_global}) by using instead a sequence of conditions $c_1, c_2, \dots, c_K$, each belonging to different parts of the sequence, to offer fine-grained control through:  
\begin{equation}
    p(x_t | x_{<t}; \, c_1, c_2, \dots, c_K ) \,.
    \label{eq:cond_local_memory}
\end{equation}
This can be implemented, for example, by treating representations of $c_1, c_2, \dots, c_K$ as additional memory for the Transformer decoder to attend to, which can be achieved with minimal modifications to \cite{li2020optimus}. 

In this paper, we particularly address the case where the target sequences can be, by nature, divided into multiple meaningful, non-overlapping segments, 
such as sentences in a text article, or bars (measures) in a musical piece.
That is to say, for sequences with $K$ segments, each timestep index $t \in [1, T]$ belongs to one of the $K$ sets of indices $I_1, I_2, \dots, I_K$, where $I_k \cap I_{k'} = \varnothing$ for $k \neq k'$ and $\bigcup_{k=1}^K I_k = [1, T]$. 
Thus, we can provide the generative model with each segment-level condition $c_k$ 
during the corresponding time interval $I_k$, leading to:
\begin{equation}
    p(x_t | x_{<t}; \, c_k ) \,, \quad \text{for}~ t \in I_k \,.
    \label{eq:cond_local_ours}
\end{equation}
This conditional generation task is different from the one outlined in Eq.~(\ref{eq:cond_local_memory}) in that we know specifically which condition (among the $K$ conditions) is in force at each timestep.
The segment-level condition $c_k$ may manifest itself as a sentence embedding in text, or a bar-level embedding in music.
In our case, we consider the conditions to be bar-level embeddings of a musical piece, i.e., we represent each $c_k$ in the vector form $\bm{c}_k \in \mathbb{R}^{d_c}$, where $d_c$ is the dimensionality of the bar-level embedding.
Collectively, the time-varying conditions $\bm{c}_1, \bm{c}_2, \dots, \bm{c}_K$ offer a \textit{blueprint}, or high-level planning, of the sequence to model. 
This is potentially helpful for generating long sequences, especially for generating music, as ancestral sampling from unconditional autoregressive models, which are trained only with long sequences of scattered tokens, may fail to exhibit the high-level flow and twists that are central to a catchy piece of music.

The \textit{segment-level conditional generation} scenario associated with Eq.~(\ref{eq:cond_local_ours}), though seemingly intuitive for long sequence modeling, has not been much studied in the literature.
Hence, we aim
to design methods that specifically tackle this task, and perform comprehensive evaluation to find out the most effective way.

\subsection{Method}\label{sec:transformers}
\begin{figure*}
    \centering
    \begin{subfigure}[b]{0.27\textwidth}
        \centering
        \includegraphics[width=\textwidth]{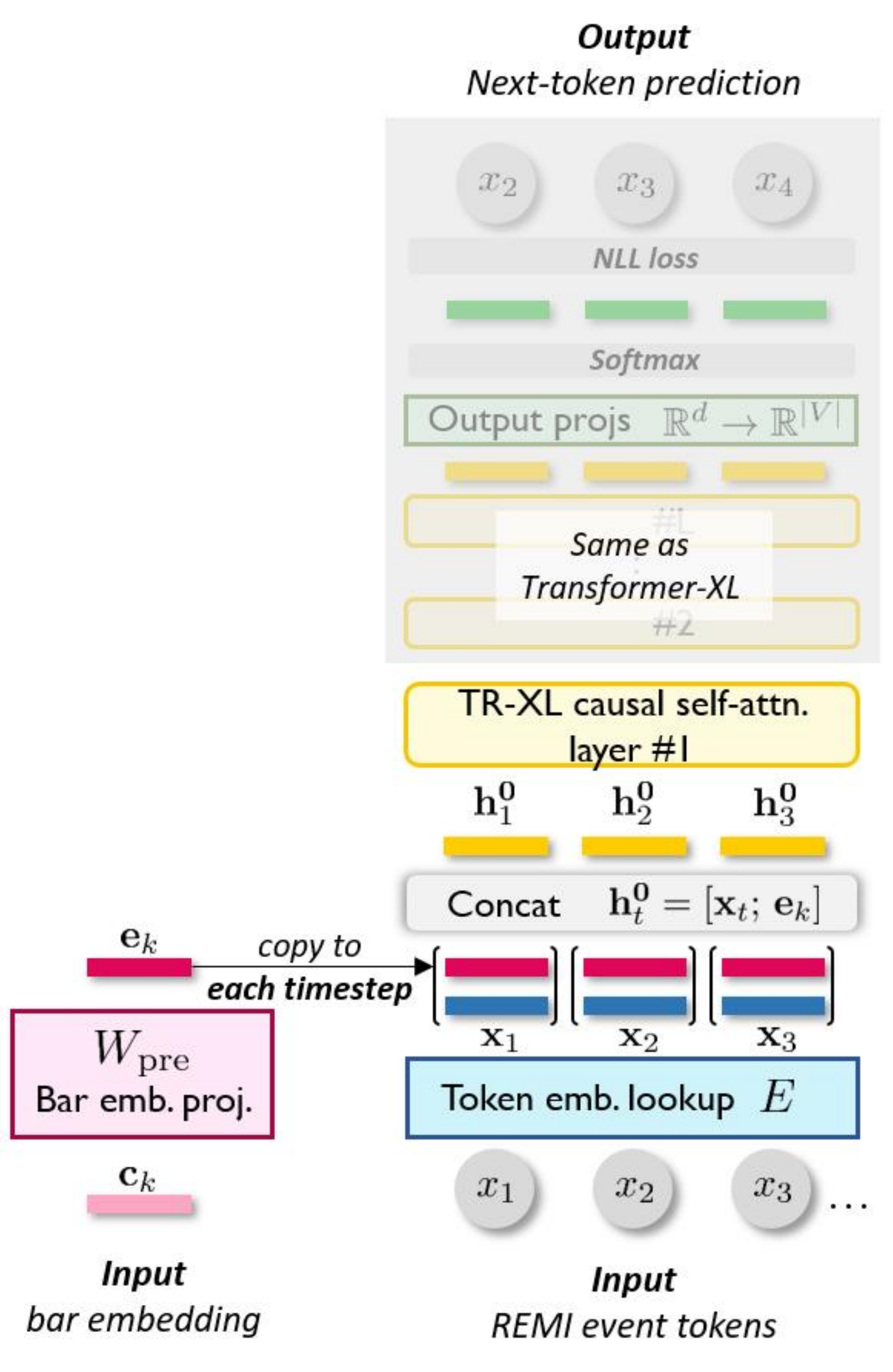}
        \caption{Pre-attention conditioning}
        \label{subfig:pre-attention}
    \end{subfigure}
    \hfill
    \centering
    \begin{subfigure}[b]{0.27\textwidth}
        \centering
        \includegraphics[width=\textwidth]{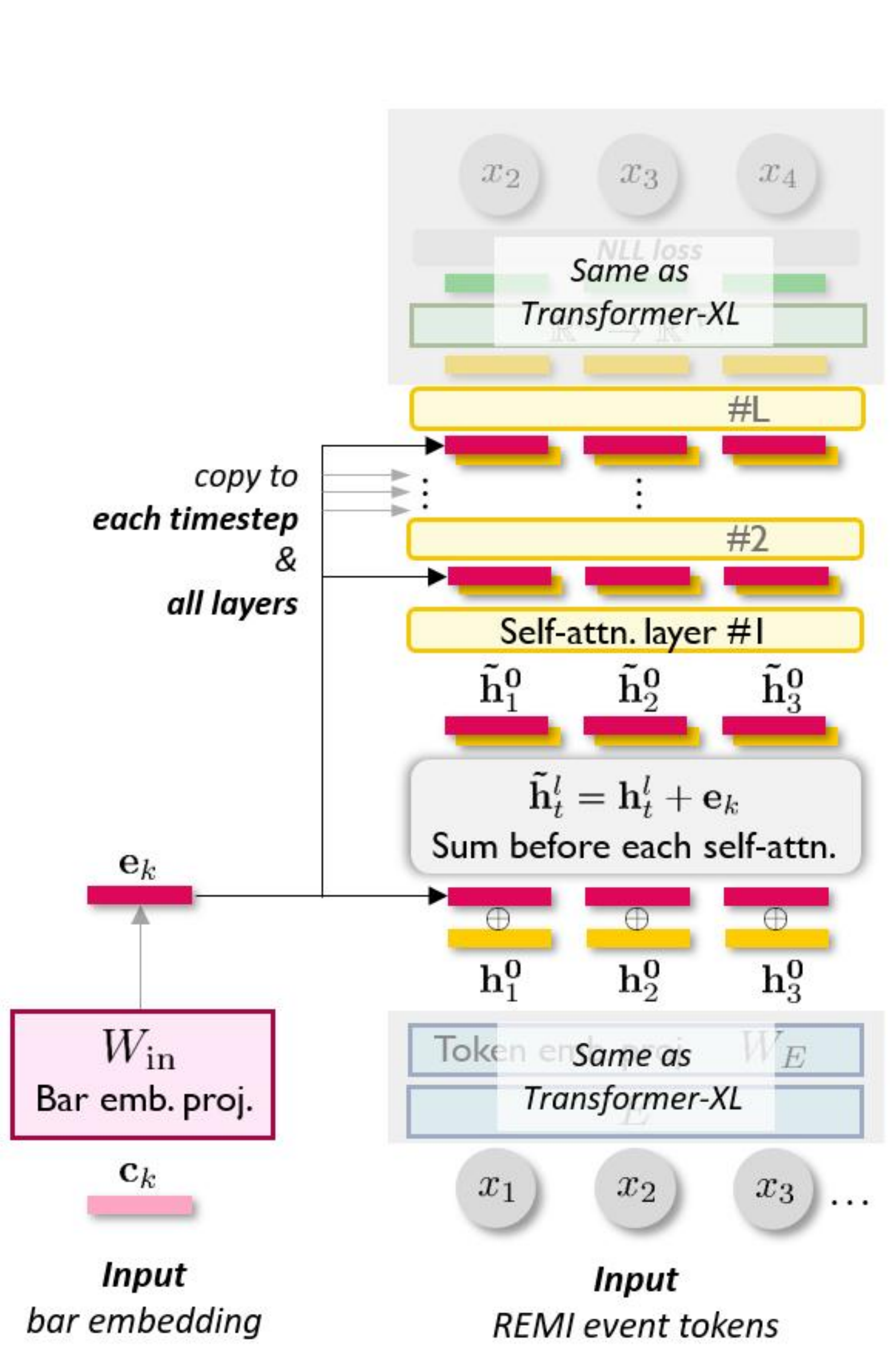}
        \caption{In-attention conditioning}
        \label{subfig:in-attention}
    \end{subfigure}
    \hfill
    \centering
    \begin{subfigure}[b]{0.287\textwidth}
        \centering
        \includegraphics[width=\textwidth]{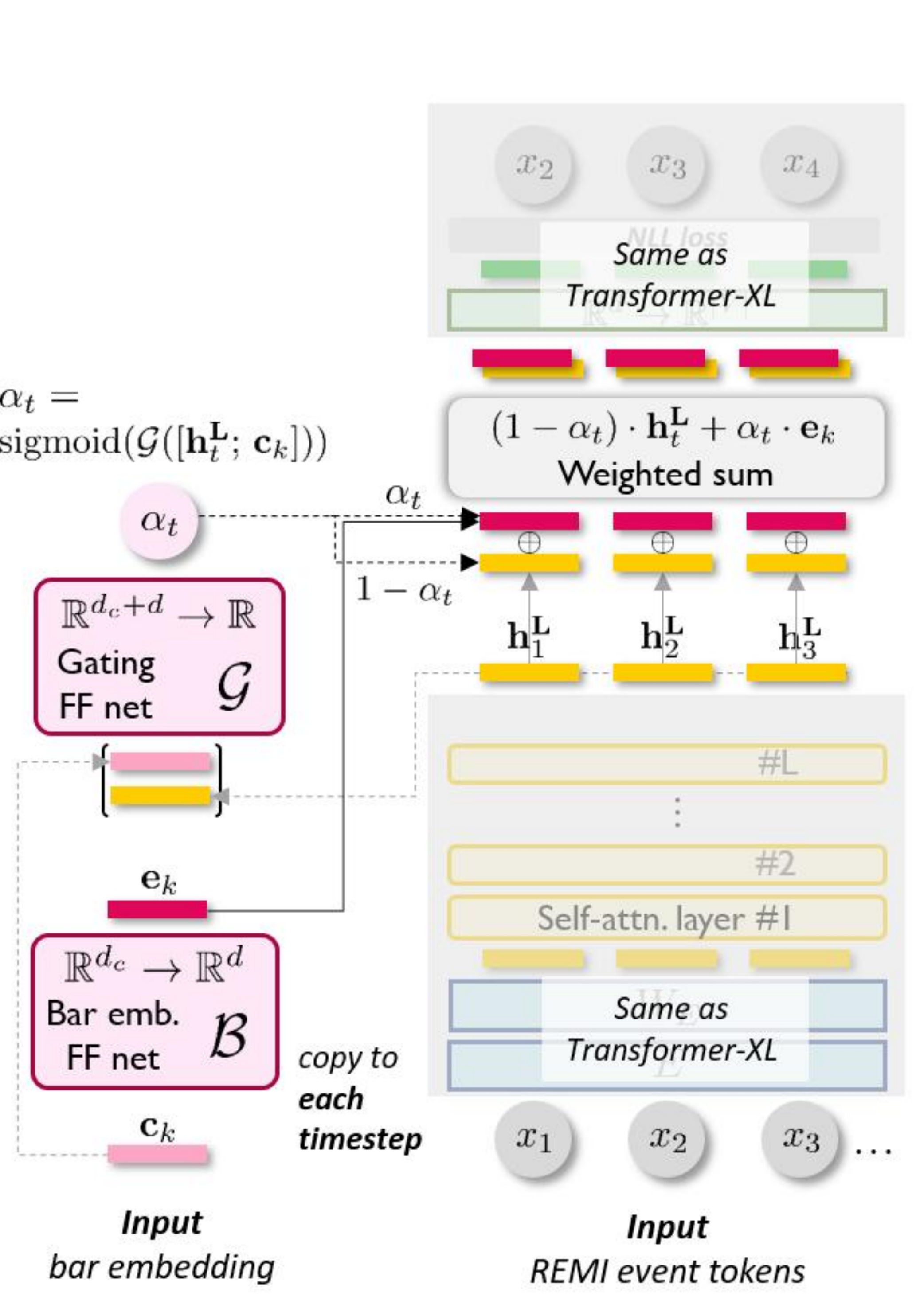}
        \caption{Post-attention conditioning}
        \label{subfig:post-attention}
    \end{subfigure}
    \caption{The architecture of segment-level conditional Transformer decoders using different conditioning mechanisms.}
    \label{fig:transformers}
\end{figure*}

Here, we elaborate the  three proposed mechanisms: \textit{pre-attention}, \textit{in-attention}, and \textit{post-attention} (see Fig.~\ref{fig:transformers} for a schematic comparison), to condition Transformer decoders at segment level.
For simplicity, we assumed here that the bar-level conditions, i.e., $\bm{c}_k$'s, can be extracted beforehand; the extraction method is also to be introduced.
Finally, we explain the data representation and dataset adopted for this task.

\textbf{Pre-attention Conditioning.}
Under pre-attention, the segment-level conditions enter the Transformer decoder only once \textit{before} all the self-attention layers. The segment embedding $\bm{c}_k$ is first transformed to a hidden condition state $\bm{e}_k$ by a matrix $W_{\text{pre}} \in \mathbb{R}^{d_\text{c} \times d_{\text{e}}}$. Then, it is concatenated with the token embedding of each timestep in the k$^{\text{th}}$ bar (i.e.,  $I_k$) before the first self-attention layer, i.e.,
\begin{equation}
    \bm{h}^0_t = \text{concat}([\bm{x}_t; \, \bm{e}_k]) \,, \quad \bm{x}_t, \bm{e}_k \in \mathbb{R}^{d_{\text{e}}} \, ,
\end{equation}
where $d_e$ is the dimensionality for token embeddings.
Note that under this mechanism, the hidden dimension, $d$, for attention modules is implicitly required to be $2d_{\text{e}}$.

The pre-attention conditioning is methodology-wise similar to the use of  segment embeddings in BERT \cite{devlin18bert}, except that BERT's segment embedding is summed directly with the token embedding, while we use concatenation to strengthen the presence of the conditions. 

\textbf{In-attention Conditioning.}
The proposed in-attention mechanism more frequently reminds the Transformer decoder of the segment-level conditions \textit{throughout} the self-attention layers. 
It projects the segment embedding $\bm{c}_k$ to the same space as the self-attention hidden states via:
\begin{equation}
    \bm{e}^{\top}_k = \bm{c}^{\top}_k W_\text{in}; \;\; W_\text{in} \in \mathbb{R}^{d_\text{c} \times d},
\end{equation}
and then sums the obtained hidden condition state $\bm{e}_k$ with the hidden states of all the self-attention layers but the last one, to form the input to the subsequent layer, i.e.,
\begin{equation}
\begin{split}
    \tilde{\bm{h}}^l_t &= \bm{h}^l_t + \bm{e}_k\,, \quad \forall l \in \{0, \dots, L-1\} \,; \\
    \bm{h}^{l+1}_t &= \text{SelfAttention}(\tilde{\bm{h}}^l_t) \,.
    \label{eq:in-attn}
\end{split}
\end{equation}
We use summation rather than concatenation here to keep the residual connections intact. We anticipate that by copy-pasting the segment-level condition everywhere, its influence on the Transformer can be further aggrandized.

\textbf{Post-attention Conditioning.}
Unlike the last two mechanisms, in post-attention, the segment embeddings do not interact with the self-attention layers at all. Instead, they are imposed \textit{afterwards} on the final attention outputs (i.e., $\bm{h}^L_t$) via two single-hidden-layer, parametric ReLU-activated feed-forward networks: the \textit{conditioning net} $\mathcal{B}$, and the \textit{gating net} $\mathcal{G}$. 
Specifically, the segment embedding $\bm{c}_k$ is transformed by the conditioning net $\mathcal{B}$ to become $\bm{e}_k \in \mathbb{R} ^d$.
Then, a gating mechanism, dictated by the gating net $\mathcal{G}$, determines ``how much'' of $\bm{e}_k$ should be blended with the self-attention output $\bm{h}^L_t$ at every timestep $t$, i.e.,
\begin{equation}
\begin{split}
    \bm{e}_k &= \mathcal{B}(\bm{c}_k) \,; \\
    \alpha_t &= \text{sigmoid}(\mathcal{G}(
        \text{concat}([\bm{h}^L_t; \, \bm{c}_k])
    )) \,;\\
    \tilde{\bm{h}}^L_t &= (1 - \alpha_t) \cdot \bm{h}^L_t + \alpha_t \cdot \bm{e}_k \,.
\end{split}
\end{equation}
Finally, $\tilde{\bm{h}}^L_t$ proceeds to the output projections
to form the probability distribution of $x_{t+1}$.

This mechanism is adapted from the \textit{contextualized vocabulary bias} proposed in the \textit{Insertion Transformer} \cite{stern2019insertion}. It is designed such that the segment-level conditions can directly bias the output event distributions, and that the model can freely decide how much it refers to the conditions for help at different timesteps within each bar.

\textbf{Implementation and Training Details.}
We adopt a 12-layer Transformer-XL \cite{dai2019transformer} as the backbone sequence model behind all of our conditioning mechanisms. To demonstrate the superiority of using Eq.~(\ref{eq:cond_local_ours}), two baselines, dubbed \textit{unconditional} and \textit{memory} hereafter, are involved in our study. 
The former takes no $\bm{c}_k$'s, thereby modeling Eq.~(\ref{eq:uncond}) exactly. The latter models Eq.~(\ref{eq:cond_local_memory}) with a conditioning mechanism largely resembling the \textit{memory} scheme introduced in \textit{Optimus} \cite{li2020optimus}, differing only in that we have multiple conditions instead of one.
In the \textit{memory} baseline, the conditions $\{\bm{c}_k\}_{k=1}^{K}$ are each transformed by matrix $W_{\text{mem}} \in \mathbb{R}^{d_c \times Ld}$ to become hidden states $\{\bm{e}_k^l \in \mathbb{R}^{d}\}_{l=1}^{L}$ unique to each layer $l$.
These states are then fed to the decoder before the training sequence, allowing all tokens (i.e., $x_t$'s) to attend to them.
Our implemented Transformer decoders have 58.7$\sim$62.6 million trainable parameters.
Some common attributes shared among the five models are listed in Table \ref{tab:model-attr} in the supplemental materials published online.\footnote{\url{slseanwu.github.io/site-musemorphose/assets/supplement.pdf}}

All these models are trained with the Adam optimizer \cite{kingma2014adam} and teacher forcing, i.e., always feeding in correct inputs rather than those sampled from previous-timestep outputs from the model itself, to minimize negative log-likelihood:
\begin{equation}
    -\sum_{t=1}^T \log p(x_t \, | \, x_{<t} \, (; \{\bm{c}_k\}_{k=1}^{K}) ),
\end{equation}
(NLL) of the training sequences ($\{\bm{c}_k\}_{k=1}^{K}$ is present only in conditional models).
Due to limited computational resources, we truncate each song to the first 32 bars, i.e., $K=32$.
We warm-up the learning rate linearly from $0$ to $2\times 10^{-4}$ in the first 500 training steps, and use cosine learning rate decay (600,000 steps) afterwards.
Trainable parameters are randomly initialized from the gaussian $\mathcal{N}(0, 0.01^2)$. Each model is trained on one NVIDIA Tesla V100 GPU (with 32GB memory) with a batch size of 4 for around 20 epochs, which requires 2 full days.

\textbf{Extraction of Bar-level Embeddings.}
We extract the conditions $c_k$ from the sequences themselves through a separate network $\mathcal{E}$, which is actually a typical 12-layer Transformer decoder. 
The network $\mathcal{E}$ takes a sequence of event tokens corresponding to a musical bar and outputs a $d_\text{c}$-dimensional embedding vector representing the bar. 
We use a GPT-2 \cite{radford2019language} like setup and train $\mathcal{E}$
for autoregressive next-token prediction, and average-pool across all  timesteps on the hidden states of a middle layer $l$ of the Transformer to obtain the segment-level condition embedding, namely,
\begin{equation}
    \bm{c}_k =  
    \text{avgpool}([\bm{h}^l_1, \dots, \bm{h}^l_{T_k}]) \,,
\end{equation}
where $\bm{h}^l_t \in \mathbb{R}^{d_\mathcal{E}} (=\mathbb{R}^{d_\text{c}})$ is the hidden state at timestep $t$ after $l$ self-attention layers, and $T_k$ is the \# of tokens in the bar. According to \cite{chen2020generative}, the hidden states in middle layers are the best contextualized representation of the input, so we set $l = 6$.

Our model $\mathcal{E}$ has 39.6 million trainable parameters.
We use all the bars associated with our training data (i.e., not limiting to the first 32 bars of the songs) for training $\mathcal{E}$, with teacher forcing and causal self-attention masking to also minimize the NLL, i.e., $- \sum_t \log p(x_t | x_{<t})$, as we do with our segment-level conditional Transformers. 

\textbf{Dataset and Data Representation.}
In our experiments, we consider modeling long sequences of symbolic music with up to 32 bars per sequence.
Specifically, our data come from the \textit{LPD-17-cleansed} dataset \cite{bmusegan}, a pop music MIDI dataset containing $>$20K songs with at most 17 instrumental tracks (e.g., piano, strings, and drums) per song.
We take the subset of 10,626 songs with time signature 4/4 (i.e., four beats per bar), and in which the piano is playing at least half of the time.
Considering only the first 32 bars of each song, our dataset contains 650 hours of music.
We reserve 4\% of the songs (i.e., 425 songs) as the validation set for objective evaluation. 

Following REMI \cite{huang2020pop},
we represent multi-track music in the form of event tokens.
Separate sets of
\textit{Note}-related tokens, i.e., \textsc{Pitch-[trk]}, \textsc{Duration-[trk]}, and \textsc{Velocity-[trk]}, are used to represent notes played by each track.
\textit{Metric}-related tokens, namely, \textsc{Bar}, \textsc{Sub-beat}, and \textsc{Tempo} are placed to represent the progression of time. 
The \textsc{Sub-beat} tokens
divide a bar into 32 possible locations for the onset (i.e., starting time) of notes, laying an explicit time grid for the Transformer.
The \textsc{Bar} token, which marks the start of a new bar, makes it easy to associate different bar-level conditions $\bm{c}_k$ to subsequences belonging to different bars (i.e., different $I_k$'s).
Our revised REMI representation leads to a vocabulary of 3,440 unique tokens.
Detailed descriptions for all types of tokens can be found in Table \ref{tab:event-token} in our online supplemental materials.

\begin{table*}
\centering
\caption{Objective evaluation results of segment-level conditional Transformers, on re-creating multi-track music ($\downarrow$\,/\,$\uparrow$: the lower/higher the score the better). }\label{tab:obj-eval}
\begin{tabular}{l  llll |ccc}
\toprule
\multirow{2}{*}{\textit{Model}} & 
\multicolumn{4}{c|}{\textbf{Fidelity}} &
\multicolumn{3}{c}{\textbf{Quality}} 
\\
& $\mathit{sim}_{\text{chr}}\uparrow$ & $\mathit{sim}_{\text{grv}}\uparrow$ & $\mathit{sim}_{\text{ins}}\uparrow$ & $\mathit{dist}_{\text{SSM}}\downarrow$
& $\mathit{KL}_\text{chr}\downarrow$ & $\mathit{KL}_\text{grv}\downarrow$ & $\mathit{KL}_\text{ins}\downarrow$  \\
\midrule
\textit{random}    & 34.6 & 40.5 & 80.4 & 13.5 & --- & --- & ---  \\
\textbf{unconditional}    & --- & --- & --- & --- & .047 & .228 & .219  \\ 
\hline
\textbf{memory} \cite{li2020optimus}   & 60.7 & 62.7 & 90.7 & 10.3 & .021 & .061 & \textbf{.006}  \\ 
\textbf{pre-attention}   & 94.5 & 92.9 & 93.0 & 7.10 & .006 & .009 & .034  \\ 
\textbf{in-attention}  & \textbf{96.3}$^{***}$ & \textbf{95.7}$^{***}$ & \textbf{97.0}$^{***}$ & \textbf{5.83}$^{***}$  & \textbf{.002} & \textbf{.005} & .021 \\ 
\textbf{post-attention}   & 92.4 & 85.7 & 95.0 & 8.27 & .016 & .054 & .056  \\
\bottomrule
\multicolumn{8}{l}{\footnotesize{$^{***}$: leads all other models with $p < .001$}} \\
\end{tabular}
\end{table*}

\subsection{Evaluation Metrics}\label{sec_eva_obj1}
To examine how well the  segment-level conditional Transformers exploit given segment-level conditions,
we let them generate 32-bar-long MIDI music, with the bar-level conditions (i.e., $\{\bm{c}_k\}_{k=1}^{32}$) extracted from songs in the validation set.
Hence, in essence, the models generate \textit{re-creations}, or \textit{covers}, of existing music.
In what follows, we define and elaborate on the objective 
metrics that assess re-creations of music.

\textbf{Evaluation Metrics for Re-creation Fidelity.}
Three bar-level metrics and a sequence-level metric are defined to quantitatively evaluate whether the re-creations of our models are similar to the original songs from which the segment-level conditions are extracted. 
\begin{itemize}[leftmargin=*]
    \item \textbf{Chroma similarity}, or $\mathit{sim}_\text{chr}$, measures the closeness of two bars in tone via:
    \begin{equation}
    \mathit{sim}_\text{chr}(\bm{r}^\text{a}, \bm{r}^\text{b}) =  100~ \frac{\langle \bm{r}^\text{a}, \bm{r}^\text{b} \rangle}{||\bm{r}^\text{a}||\,||\bm{r}^\text{b}||},
    \end{equation}

    where $\langle \cdot,\cdot\rangle$ denotes dot-product, and $\bm{r}\in \mathbb{Z}^{12}$ is the \emph{chroma vector}  \cite{fujishima99} representing the number of onsets for each of the 12 pitch classes (i.e., \textsc{C}, \textsc{C\#}, \dots, \textsc{B}) within a bar, counted across octaves and tracks, with the drums track ignored. 
    One of the bars in the pair $(\text{a}, \text{b})$ is from the original song, while the other is from the re-creation.

\item \textbf{Grooving similarity}, or $\mathit{sim}_\text{grv}$,  examines the rhythmic resemblance between two bars with the same formulation as the chroma similarity, but measured on the \textit{grooving vectors} $\bm{g} \in \mathbb{Z}^{32}$ recording the number of note onsets, counted across tracks, that occur at each of the 32 sub-beats in a bar \cite{dixonEtAl04ismir}. 

\item \textbf{Instrumentation similarity}, or $\mathit{sim}_\text{ins}$, quantifies whether two bars use similar instruments via:
\begin{equation}
    \mathit{sim}_\text{ins}(\bm{b}^\text{a}, \bm{b}^\text{b}) = 100~ (1 - \frac{1}{17} \sum_{i=1}^{17} \text{XOR}(b_i^\text{a}, b_i^\text{b}) ),
\end{equation}
where 
$\bm{b} \in \{0, 1\}^{17}$ is a binary vector indicating
the presence of a track in a bar, i.e., at least one \textsc{Pitch-[trk]} for that \textsc{[trk]} occurs within the bar.

\item \textbf{Self-similarity matrix distance}, or $\mathit{dist_\text{SSM}}$, measures whether two 32-bar sequences have similar overall structure by comparing the mean absolute distance between the \textit{self-similarity matrices} (SSM) \cite{foote1999visualizing} of the two sequences, i.e.,
\begin{equation}
    100\,||S^\text{a}-S^\text{b}||_1,
\end{equation} 
where $S \in \mathbb{R}^{N \times N}$ is the SSM. 
In doing so, we firstly synthesize each 32-bar sequence into audio with a synthesizer. Then, we use constant-Q transform to extract acoustic chroma vectors indicating the relative strength of each pitch class for each half beat in an audio piece, and then calculate the cosine similarity among such \textit{beat-synchronized}  features; hence, $s_{ij} \in [0, 1]$ for half beats $i$ and $j$ within the same music piece. Since each bar in 4/4 signature has 8 half-beats, we have $N=256$ for our data. 
An SSM depicts the self-repetition within a piece.
Unlike the previous metrics, the value of $\mathit{dist_\text{SSM}}$ is the closer to zero the better. 
We note that the values of all metrics above, i.e., $\mathit{sim}_\text{chr}, \mathit{sim}_\text{grv}$, $\mathit{sim}_\text{ins}$, and $\mathit{dist_\text{SSM}}$, are all in $[0,100]$. 
\end{itemize}

\textbf{Evaluation Metrics for Re-creation Quality.}
It is worth mentioning that objective evaluation on the quality of generated music remains an open issue \cite{wu2020jazz}. 
Nevertheless, we adopt here a recent idea that measures the quality of music by comparing the \textit{distributions} of the generated content and the training data in feature spaces, using KL divergence \cite{yang20evaluation}.

While we employ $\mathit{sim}_\text{chr}$, $\mathit{sim}_\text{grv}$, $\mathit{sim}_\text{ins}$ to measure the similarity between a generated bar and the corresponding reference bar in Sec.~\ref{sec_eva_obj1}, 
here we instead use them to calculate either the  (self-)similarity among the bars in a \emph{gen}erated 32-bar sequence, or that among the bars in a \emph{real} training sequence. For each sequence, we have $32\cdot(32-1)/2$ such pairs, leading to a distribution. 
We discretize the distributions of $\mathit{sim}_{\text{chr}}$ and $\mathit{sim}_{\text{grv}}$ into 50 bins with evenly-spaced boundaries $[0, 2, \dots, 100]$. 
The KL divergence is calculated on the probability mass functions $p_\text{real}, p_\text{gen} \in \mathbb{R}^{50}$, or $\mathbb{R}^{18}$ for the distribution in terms of $\mathit{sim}_{\text{ins}}$,  and is defined as $\mathit{KL}\, (p_\text{real}\, || \, p_\text{gen})$.
We assume that the values are the closer to zero the better.

\begin{figure}
    \centering
    \begin{subfigure}[b]{0.45\linewidth}
        \centering
        \includegraphics[width=\textwidth]{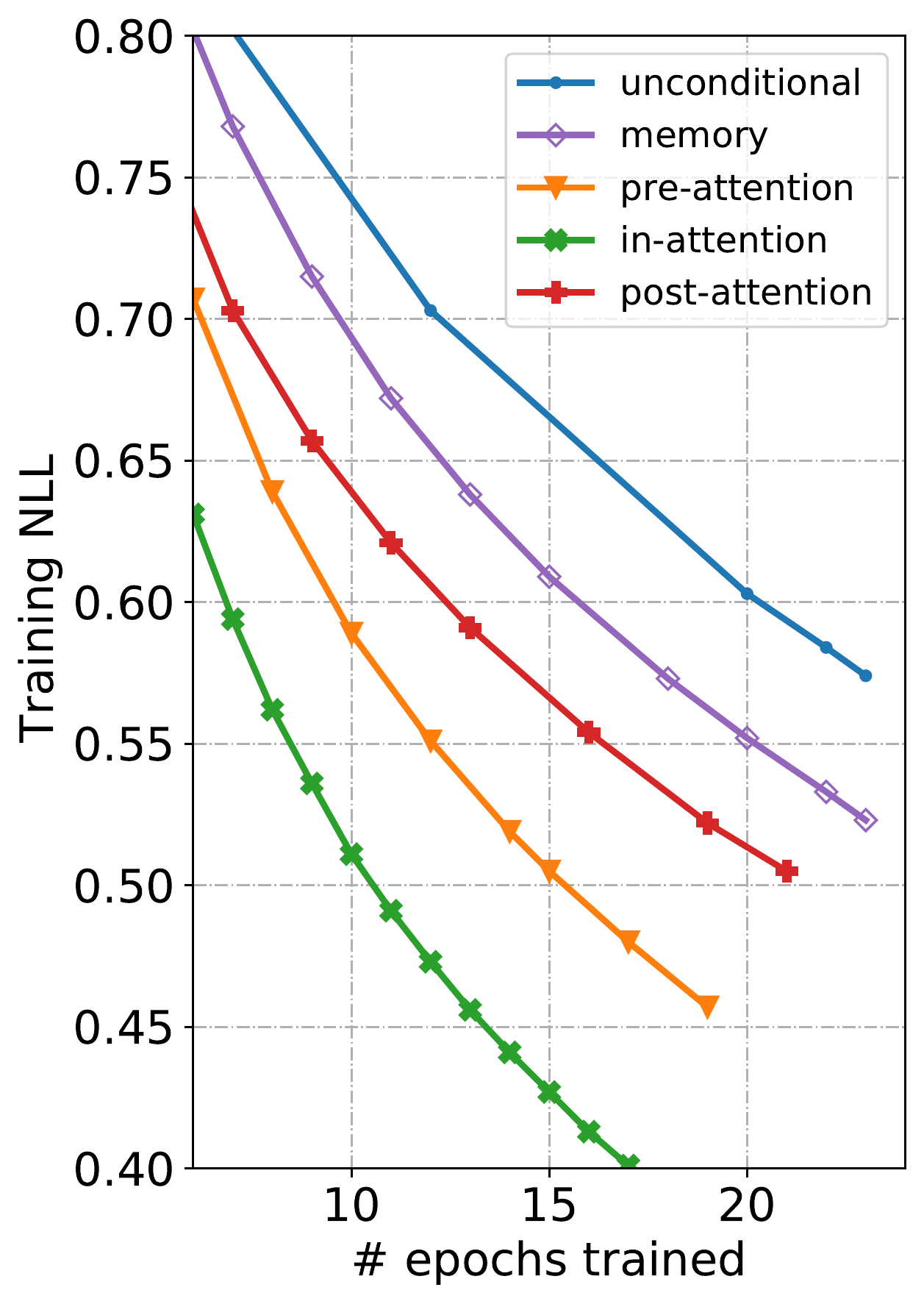}
        \caption{\# epochs vs. training loss}
        \label{subfig:train-prgs}
    \end{subfigure}
    \hfill
    \centering
    \begin{subfigure}[b]{0.465\linewidth}
        \centering
        \includegraphics[width=\textwidth]{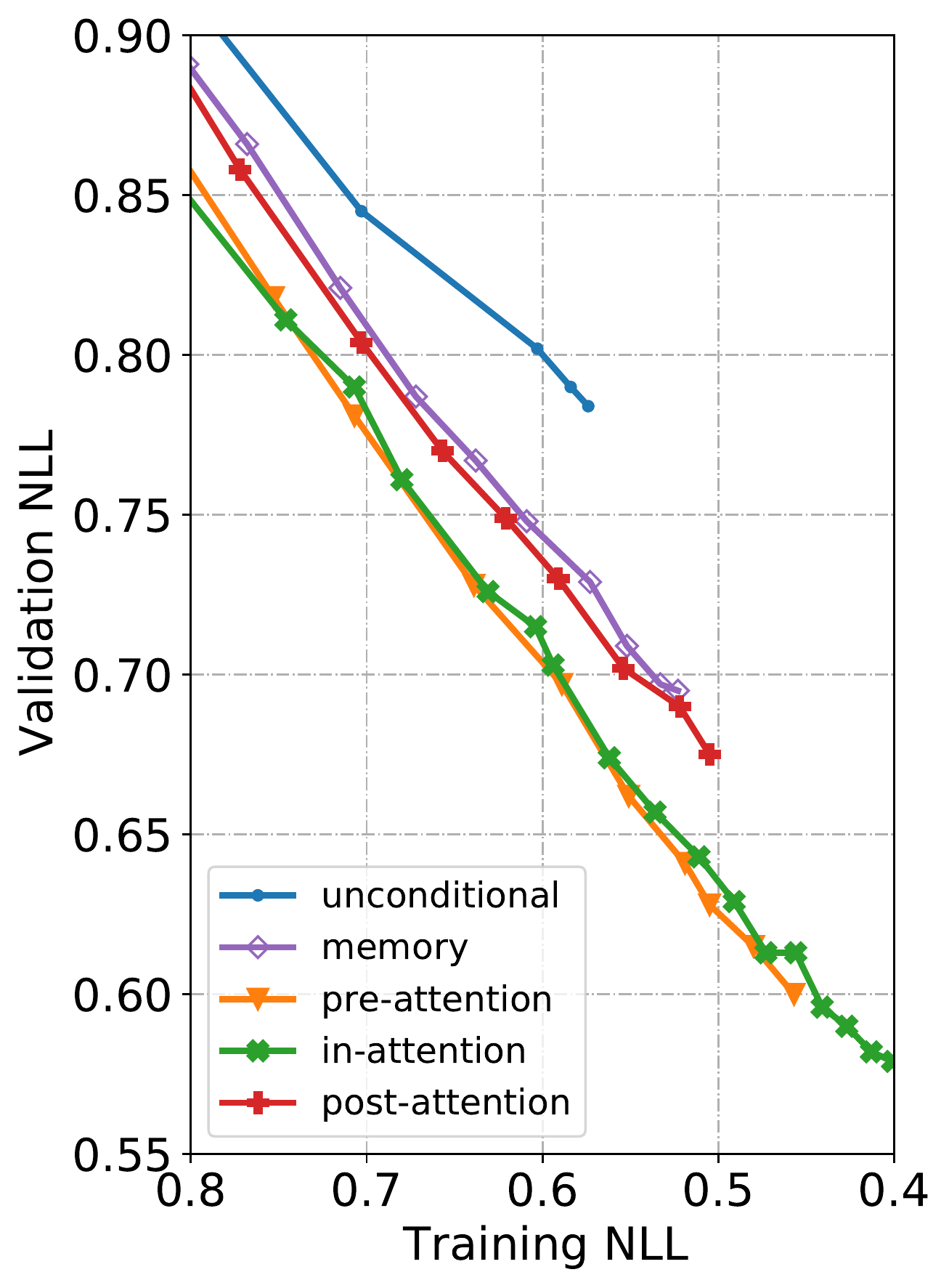}
        \caption{Training loss vs. val. loss}
        \label{subfig:valloss}
    \end{subfigure}
    \caption{Training dynamics of segment-level conditional Transformers on \textit{LPD-17-cleansed} dataset (view in color).}
    \label{fig:train-prgs}
\end{figure}

\subsection{Results and Discussion}
\begin{figure}
    \centering
    \begin{subfigure}[b]{0.43\linewidth}
        \centering
        \includegraphics[width=\textwidth]{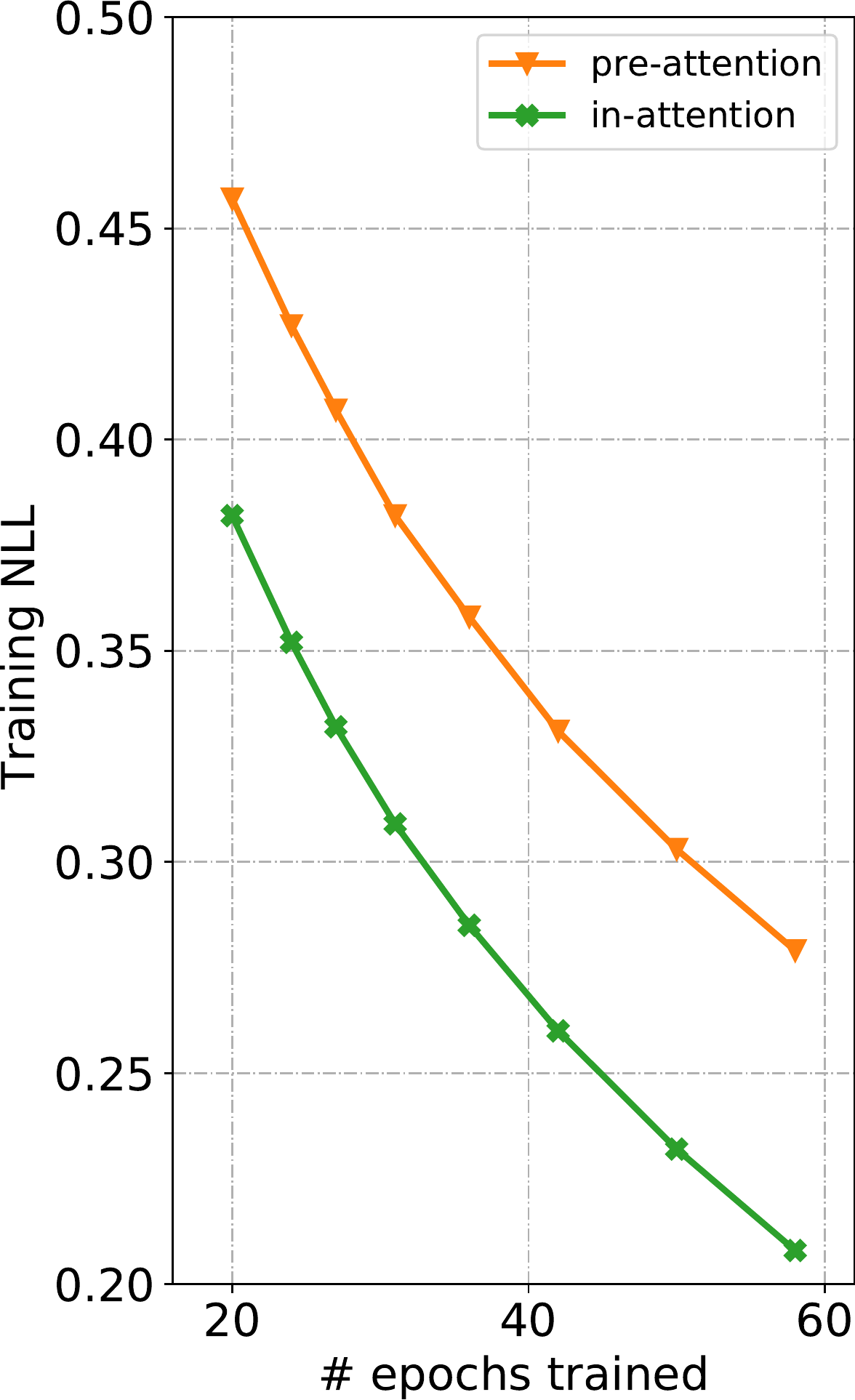}
        \caption{\# epochs vs. training loss}
        \label{subfig:train-prgs-cont}
    \end{subfigure}
    \hfill
    \centering
    \begin{subfigure}[b]{0.515\linewidth}
        \centering
        \includegraphics[width=\textwidth]{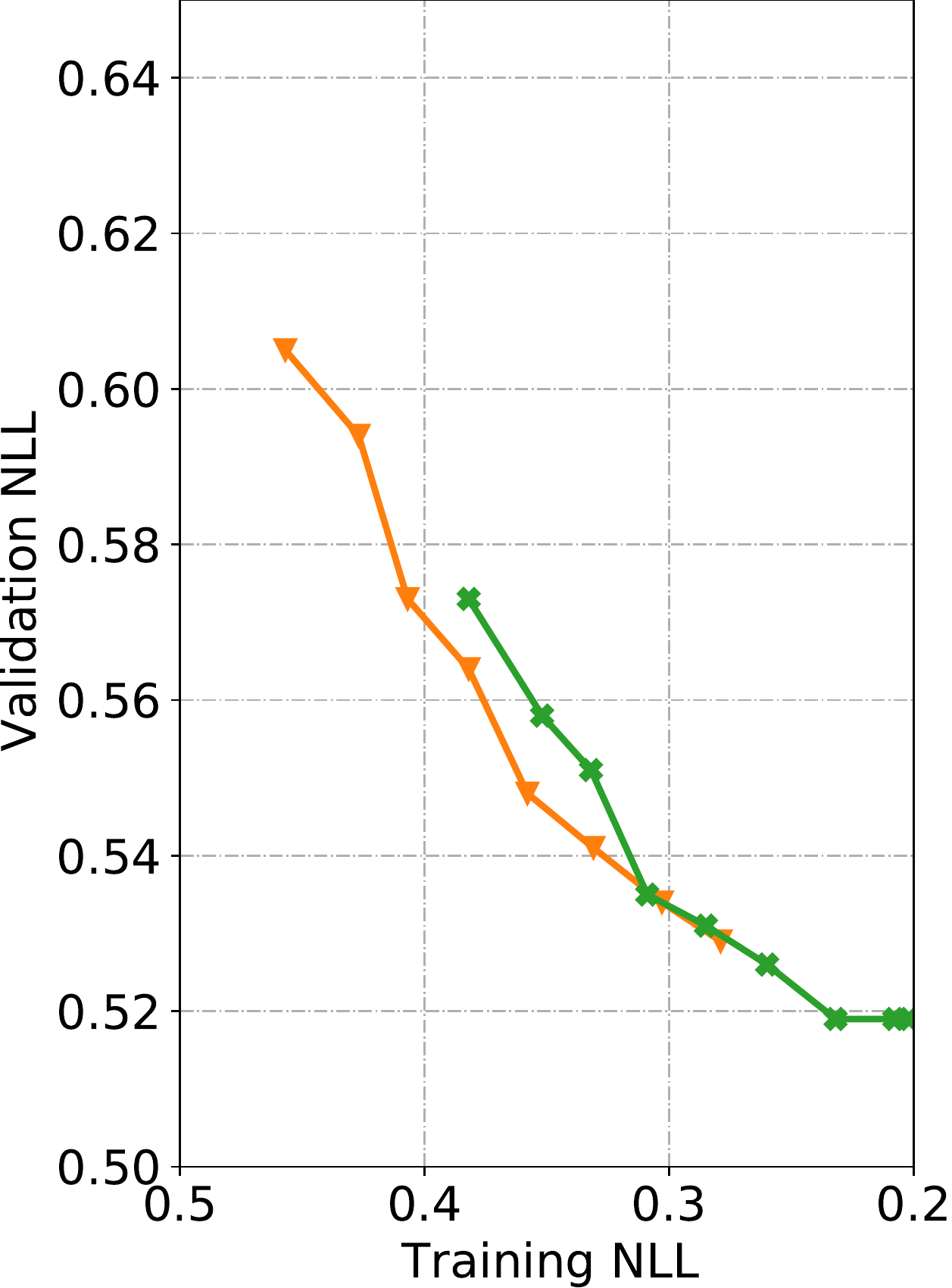}
        \caption{Training loss vs. val. loss}
        \label{subfig:valloss-cont}
    \end{subfigure}
    \caption{Training dynamics of \textit{pre-attention} and \textit{in-attention} Transformers on \textit{LPD-17-cleansed} dataset from the 20th epoch onwards (best viewed in color).}
    \label{fig:train-prgs-cont}
\end{figure}

We begin with presenting the training dynamics of our models (see Fig.~\ref{fig:train-prgs}).
From Fig.~\ref{subfig:train-prgs}, we can see that the model whose loss drops the fastest, \textit{in-attention}, achieves 0.41 training NLL after 16 epochs, while the \textit{unconditional} and \textit{memory} \cite{li2020optimus} models still have a loss of around 0.65 and 0.60 respectively. 
We also take validation loss into consideration. Fig.~\ref{subfig:valloss} 
shows that our models, especially \textit{pre-} and \textit{in-attention}, outperform both \textit{unconditional} and \textit{memory} at comparable training NLL.
\revision{Given the overlapping lines of \textit{pre-attention} and \textit{in-attention} on Fig.~\ref{subfig:valloss}, some readers may be interested in how they compare against each other with more training epochs.
We plot the comparison (until the 58th epoch) in Fig.~\ref{fig:train-prgs-cont} in the online supplemental.
We find that \textit{pre-attention}'s validation NLL slowly catches up with that of \textit{in-attention}, but has not quite match it after an additional week of training.\footnote{\revision{In Sec.~\ref{sec:part2}, we will show that \textit{pre-attention} performs far worse than \textit{in-attention} when paired with a VAE training objective.}}}

Table \ref{tab:obj-eval} displays the objective evaluation results.
The scores are calculated on 425 re-creations, each using the bar-level conditions from a unique song in the validation set.
For comparison, we let the unconditional model randomly generate 400 32-bar pieces from scratch,
and further include a \textit{random} baseline, whose scores are computed over 400 random pairs of pieces drawn from the training set.

Focusing on the fidelity metrics first, all of our conditional models score quite high on bar-level $\mathit{sim}_\text{chr}, \mathit{sim}_\text{grv}$, and $\mathit{sim}_\text{ins}$, with \textit{in-attention} significantly outperforming the rest.
\textit{In-attention} also attains the lowest $\mathit{dist}_{\text{SSM}}$, 
demonstrating effective and tight control.
Moreover, it is worth noting that the \textit{memory} model sits right in between ours and the \textit{random} baseline in all metrics, offering limited conditioning ability.

Next, we shift attention to the quality metrics,
where $\mathit{KL}_\text{chr}$, $\mathit{KL}_\text{grv}$, and $\mathit{KL}_\text{ins}$ represent the KL divergence calculated on the corresponding distributions.
The conditional models clearly outperform the unconditional baseline, suggesting that our conditions do help Transformers in modeling music.
Furthermore, once again, \textit{in-attention} comes out on top, followed by \textit{pre-}, \textit{post-attention}, and then \textit{memory}, except for metric $\mathit{KL}_\text{ins}$, in which \textit{memory} performs the best.

From the results presented above, we surmise that the generally worse performance of \textit{memory} baseline compared to ours could be due to the advantages of using Eq.~(\ref{eq:cond_local_ours}) over Eq.~(\ref{eq:cond_local_memory}). 
Equation (\ref{eq:cond_local_ours}) informs the Transformer decoder of when exactly to exploit each of the conditions, and eliminates the bias caused by the positional encoding fed to Transformers, i.e., later tokens in the sequence are made more dissimilar to the conditions provided in the beginning of the sequence, thereby undermining the conditions' effect in the attention process. 
Therefore, \textit{memory} works well only for conditions that remain relatively unchanged throughout a sequence, such as the instrumentation of a song.  
Among the three proposed conditioning methods, we conjecture that \textit{post-attention} works the worst because the conditions do not participate in the attention process. Therefore, the model has less chance to integrate information from them. Moreover, \textit{in-attention} possesses an advantage over \textit{pre-attention} reasonably because it feeds conditions to all attention layers instead of just once before the first attention layer.
With the comprehensive evaluation conducted, we have sufficient evidence indicating that \textit{in-attention} works the best in exerting control over Transformer decoders with segment-level, time-varying conditions.

\section{MuseMorphose: Generating Music with Time-Varying User Controls}
\label{sec:part2}

We have developed in Section \ref{sec:part1} the \textit{in-attention} technique, which can exert firm control over a Transformer decoder's generation with time-varying, predetermined conditioning vectors.
However, that technique alone only enables Transformers to compose re-creations of music
at random.
No freedom has been given to users to interact with such a system to affect the music it generates as one wishes,
thereby limiting its practical value.
This section aims at alleviating such a limitation.
We bridge the \textit{in-attention} Transformer decoder and a jointly learned bar-wise Transformer encoder,
which is tasked with learning the bar-level latent conditions.
We train the encoder-decoder network with the variational autoencoder (VAE) training objective, 
and introduce \textit{attribute embeddings} \cite{fu2018style, kawai2020attributes}, also working at the bar level, to the decoder to discourage the latent conditions from storing attribute-related information and realize fine-grained user controls over sequence generation.
Experiments show that our resulting model, \textit{MuseMorphose}, successfully allows users to harness two easily perceptible musical attributes, \textit{rhythmic intensity} and \textit{polyphony}, while maintaining fluency and high fidelity to the input reference song. 
Such combination of abilities is shown unattainable by existing methods \cite{brunner2018midi, kawai2020attributes}.

\subsection{Problem Formulation}
For clarity, we model our user-controllable conditional generation task as a style transfer problem without paired data \cite{fu2018style}.
This problem, in its most basic form, deals with an input instance $X$, such as a sentence or a musical excerpt, and an attribute $a$ intrinsic to $X$. 
The attribute $a \in \{1,\dots,C\}$ can be either a nominal (e.g., composer style), or an ordinal (e.g., note density) variable with $C$ categories.
As users, we would like to have a model that takes a tuple:
\begin{equation}
    (X, \tilde{a}),
\end{equation}
where $\tilde{a}$ is the user-specified attribute value that may be different from the original $a$, and outputs a style-transferred instance $\tilde{X}$.
A desirable outcome is that the generated $\tilde{X}$ bears the specified attribute value $\tilde{a}$ (i.e., achieving effective \textit{control}), and preserves the content other than the attribute (i.e., exhibiting high \textit{fidelity}). 
Being able to generate \textit{diverse} versions of  $\tilde{X}$'s given the same $\tilde{a}$ or to ensure the \textit{fluency} of generations, though not central to this problem, are also preferred characteristics of such models.
This framework has been widely adopted by past works in both text \cite{fu2018style, yang2018unsupervised, john2019disentangled} and music \cite{brunner2018midi} domains.

A natural extension to the problem above is to augment the model such that it can process multiple attributes at a time, i.e., taking inputs in the form:
\begin{equation}
    (X, \tilde{a}^1, \tilde{a}^2,\dots,\tilde{a}^J),
\end{equation}
to produce $\tilde{X}$, where ${a}^j \text{, for} \; j\in\{1, J\}$, are the $J$ categorical attributes being considered.
In this case, users have the freedom to alter single or multiple attributes. Hence, preferably, changing one attribute to $\tilde{a}^j$ should not affect an unaltered attribute $a^{j'}$.
This multi-attribute variant has also been studied in \cite{lample2018multiple, kawai2020attributes}.

In some scenarios, the input sequence $X$ is long and can be partitioned into meaningful continuous segments, e.g., sentences in text or bars in music, through:
\begin{equation}
\begin{split}
    X &= \{X_1, \dots, X_K\}, \text{where} \\
    X_k &= \{x_t \, \vert \, t \in I_k  \}, \; I_k \subset \mathbb{N} \quad \text{for}~ 1\leq k \leq K; \\
    \bigcup_{k=1}^K I_k &= [1, T] \quad \text{and} \quad
    I_k \cap I_{k'} = \varnothing \; \text{for} \; k \neq k' \,, \label{eq:style-trans-part}
\end{split}
\end{equation}
where $T$ is the length of $X$, and $I_1,\dots,I_K$ constitute a partition of $X$ (cf. the paragraph about the formulation of Eq.~(\ref{eq:cond_local_ours}) in Sec.~\ref{sec:form-ch3}). 
For such cases, it could be a privilege if we can control \textit{each} segment individually, since this allows, for example, users to make a musical section a lot more intense, and another much calmer, to achieve dramatic contrast.
Therefore, we formulate here the style transfer problem with \textit{multiple}, \textit{time-varying} attributes.
A model for this problem accepts:
\begin{equation}
    (X, 
    \{\tilde{a}_k^1, \tilde{a}_k^2,\dots,\tilde{a}_k^J\}_{k=1}^K),
\end{equation}
and then generates a long, style-transferred $\tilde{X}$, where $a_k^1, a_k^2,\dots,a_k^J$ are concerned specifically with the segment $X_k$.
Though Transformers \cite{vaswani2017attention, radford2019language} are impressive in sequence generation, no prior work, to our knowledge, has addressed the problem above with them, reasonably because there has not been a well-known segment-level conditioning mechanism for Transformers.
Thus, given our developed \textit{in-attention} method, we are indeed in a privileged position to approach this task.

\subsection{Method}\label{sec:attr-comp}
In what follows, we describe how we pair an \textbf{in-attention} Transformer decoder with a bar-wise Transformer encoder, as well as attribute embeddings, to achieve fine-grained controllable music generation through optimizing the VAE objective.
The musical attributes considered here are: \textit{rhythmic intensity} ($a^\text{rhym}$) and \textit{polyphony} ($a^\text{poly}$; i.e., harmonic fullness),
whose calculation methods will also be detailed.
After that, we present the implementation details and the pop piano dataset, \textit{AILabs.tw-Pop1K7}, used for this task.

\begin{figure*}
    \centering
    \includegraphics[width=0.63\textwidth]{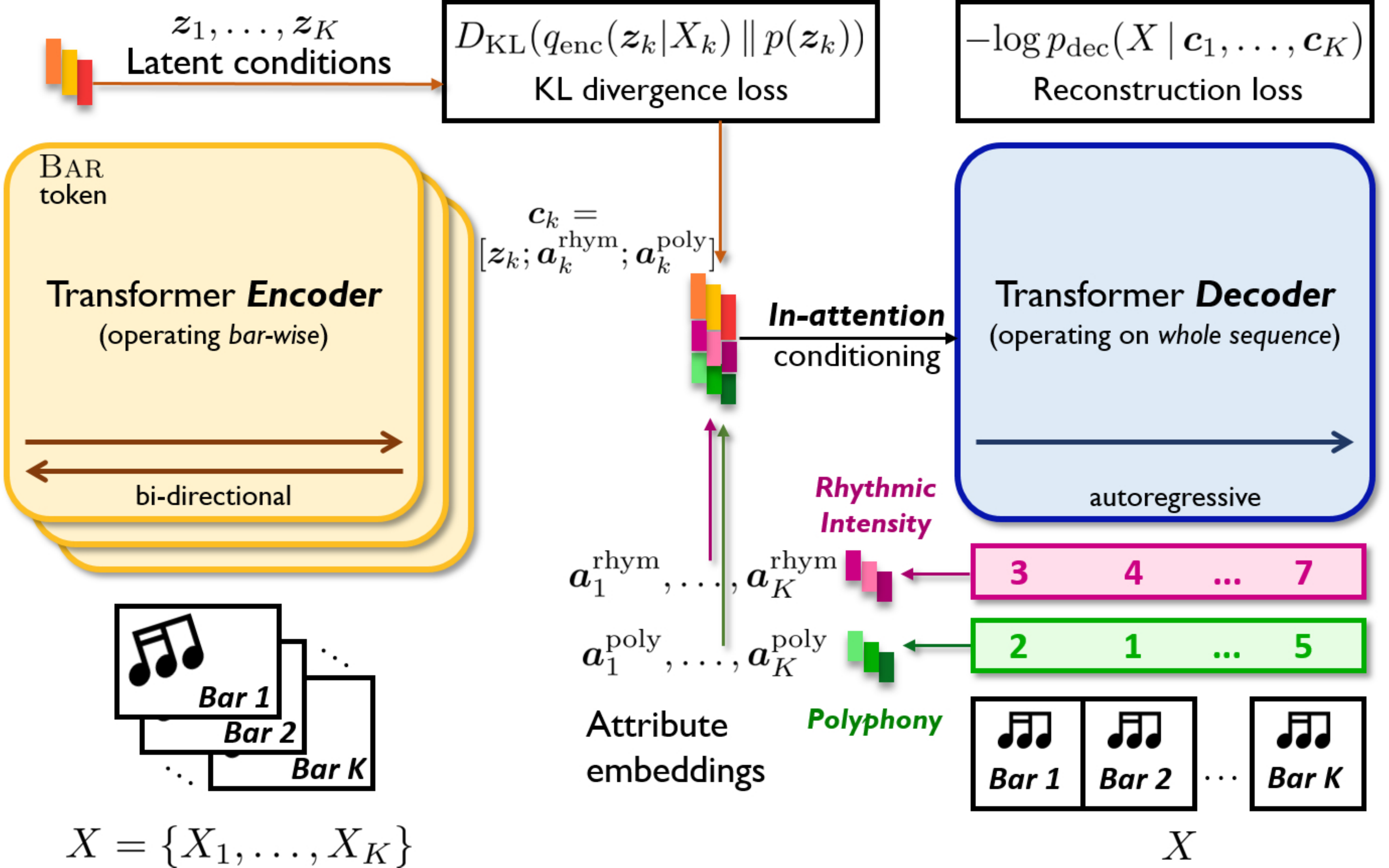}
    \vspace{1mm}
    \caption{Architecture of \textit{MuseMorphose}, a Transformer-based VAE model for fine-grained music style transfer.}
    \label{fig:muse-optimus}
\end{figure*}

\textbf{Model Architecture.}
Figure \ref{fig:muse-optimus} illustrates our \textit{MuseMorphose} model,
which consists of a vanilla Transformer encoder (i.e., the version proposed by Vaswani et al.~\cite{vaswani2017attention}) operating at bar level, a Transformer decoder that accepts segment-level conditions through the \textit{in-attention} mechanism, and a KL divergence-regularized latent space for the representation of \textit{musical bars} between them.
In MuseMorphose, the input music is partitioned as $X = \{X_1,\dots,X_K\}$, where $X_k$ houses the $k^\text{th}$ bar of the music.
The encoder works on the musical bars, $X_1,\dots,X_K$, in parallel, while the decoder sees the entire piece $X$ at once.
The bar-level attributes, $a_k^\text{rhym}$ and $a_k^\text{poly}$, each consisting of 8 ordinal classes, are transformed into embedding vectors, $\bm{a}_k^{\text{rhym}}, \bm{a}_k^{\text{poly}} \in \mathbb{R}^{d_{\bm{a}}}$, before entering the decoder through in-attention.
On the whole, the operations taking place inside MuseMorphose can be summarized as follows:
\begin{equation}
\begin{split}
    \bm{z}_k &= \textbf{enc}(X_k) \quad \text{for}~ 1\leq k \leq K; \\
    \bm{c}_k &= \text{concat}([\bm{z}_k; \bm{a}_k^{\text{rhym}}; \bm{a}_k^{\text{poly}}]); \\
    \bm{y}_{t} &= \textbf{dec}(x_{<t}; \bm{c}_k), \; t \in I_k \quad \text{for}~ 1\leq k \leq K,
\end{split}
\end{equation}
where $\bm{z}_k \in \mathbb{R}^{d_{\bm{z}}}$ is the latent condition for the $k^\text{th}$ bar;
$\bm{a}_k^{\text{rhym}}, \bm{a}_k^{\text{poly}}$ are the bar's attribute embeddings,
$I_k$ stores the timestep indices (see Eq.~(\ref{eq:style-trans-part})) of the bar; and,
$\bm{y}_{t}$ is the predicted probability distribution for $x_t$.
Note that though the input is split into bars on the encoder side, the decoder still deals with the entire sequence, thanks to both in-attention conditioning and $\bigcup_{k=1}^K I_k = [1, T]$.
This asymmetric architecture is likely advantageous since
it enables fine-grained conditions to be given, 
and also promotes the coherence of long generations.

Since we adopt the VAE framework here, we now elaborate more on how we construct the latent space for $\bm{z}_k$'s.
First, we treat the encoder's attention output at the first timestep (corresponding to the \textsc{Bar} token), i.e., $\bm{h}^{L_{\textbf{enc}}}_{k, 1}$, as the contextualized representation of the bar.
Then, following the conventional VAE setting \cite{kingma2014auto}, it is projected by two separate learnable weights $W_{\bm{\mu}}, W_{\bm{\sigma}}$, to the mean and std vectors:
\begin{equation}
    \bm{\mu}_k = {\bm{h}^{L_{\textbf{enc}}}_{k, 1}}^\top W_{\bm{\mu}} \quad \quad
    \bm{\sigma}_k = {\bm{h}^{L_{\textbf{enc}}}_{k, 1}}^\top W_{\bm{\sigma}}\,,
\end{equation}
where  $W_{\bm{\mu}}, W_{\bm{\sigma}} \in \mathbb{R}^{d \times d_{\bm{z}}}$.
Afterwards, we may sample the latent condition to be fed to the decoder from the isotropic gaussian defined by $\bm{\mu}_k$ and $\bm{\sigma}_k$:
\begin{equation}
   \bm{z}_k \sim \mathcal{N}(\bm{\mu}_k, \text{diag}(\bm{\sigma}^2_k)).
\end{equation}
For simplicity, the prior of latent conditions, i.e., $p(\bm{z}_k)$, is set to the typically-used standard gaussian $\mathcal{N}(\bm{0}, I_{d_{\bm{z}}})$.

It is worthwhile to mention that the subsequent copy-paste of the $\bm{z}_k$'s, done by the \textit{in-attention} decoder, to every attention layer and every timestep in $I_k$ coincides with the design of \textit{Skip-VAE} \cite{dieng2019avoiding},
which provably mitigates the posterior collapse problem, i.e., the case where the decoder completely ignores $\bm{z}_k$'s and degenerates into autoregressive sequence modeling.
Also, our work makes a notable advance over \textit{Optimus} \cite{li2020optimus} in that we equip Transformer-based VAEs with the capability to model long sequences (with length in the order of ${\geq}10^3$), under fine-grained changing conditions from the latent vector (i.e., $\bm{z}_k$)  and user controls (i.e., $\tilde{a}_k^\text{rhym}$ and $\tilde{a}_k^\text{poly}$).

\textbf{Training Objective.}
MuseMorphose is trained with the $\beta$-VAE objective \cite{higgins2017beta} with free bits \cite{kingma2016improving}. It minimizes the loss, $\mathcal{L}_{\text{MuseMorphose}}$, which is written as:
\begin{equation}
\begin{split}
    \mathcal{L}_{\text{MuseMorphose}} &= \mathcal{L}_{\text{NLL}} + \beta \mathcal{L}_{\text{KL}} \,, \quad \text{where} \\
    \mathcal{L}_{\text{NLL}} &= -\log p_{\textbf{dec}}(X \, \vert \, \bm{c}_1,\dots,\bm{c}_K)\,; \\
    \mathcal{L}_{\text{KL}} &= \frac{1}{K} \sum_{k=1}^K  \sum_{i=1}^{d_{\bm{z}}} \max(\lambda,   D_\text{KL}(q_{\textbf{enc}}(z_{k, i} \vert X_k) \, \| \, p(z_{k, i})))\,.\label{eq:mopt-loss}
\end{split}
\end{equation}
The first term, $\mathcal{L}_{\text{NLL}}$, is the conditional negative log-likelihood (NLL) for the decoder to generate input $X$ given the conditions $\bm{c}_1,\dots,\bm{c}_K$,
hence referred to as \textit{reconstruction NLL} hereafter.
The second term, $\mathcal{L}_{\text{KL}}$, is the KL divergence between the posterior distributions of $\bm{z_k}$'s estimated by the encoder (i.e., $q_{\textbf{enc}}(\bm{z}_{k} \vert X_k)$) and the prior $p(\bm{z}_k)$. 
$\beta$ \cite{higgins2017beta} and $\lambda$ \cite{kingma2016improving} are  hyperparameters to be tuned (see Section \ref{sec:vae-trick} for explanations). 
We refer readers to Fig.~\ref{fig:muse-optimus} for a big picture of where the two loss terms act on the model.

Previous works that employ autoencoders for style transfer tasks \cite{john2019disentangled, kawai2020attributes, dai2019style} often suggested adding adversarial losses \cite{goodfellow2014generative} on the latent space, so as to discourage it from storing style-related, or attribute-related, information.
However, a potential downside of this practice is that it introduces additional complications to the training of our Transformer-based network, which is already very complex itself.
We instead demonstrate that by using suitable $\beta$ and $\lambda$ to control the size of latent information bottleneck, both strong style transfer and good content preservation of input $X$ can be accomplished without auxiliary training losses.

\textbf{Calculation of Musical Attributes.}
The attributes chosen for our task, \textit{rhythmic intensity} ($a^\text{rhym}$) and \textit{polyphony} ($a^\text{poly}$), are able to be perceived easily by people without intensive training in music. Meanwhile,
they are also important determining factors of musical emotion.
To obtain the ordinal classes $a^\text{rhym}$ and $a^\text{poly}$ of each bar, we first compute the raw scores $s^\text{rhym}$ and $s^\text{poly}$.
\begin{itemize}[leftmargin=*]
    \item \textbf{Rhythmic intensity score}, or $s^\text{rhym}$, simply measures the percentange of sub-beats with at least one note onset, i.e.:
    \begin{equation}\label{eq:rhym-int}
        s^\text{rhym} = \frac{1}{B} \sum_{b=1}^B \bm{1}( n_{\text{onset}, b} \geq 1 ) \,,
    \end{equation}
    where $B$ is the number of sub-beats in a bar and $\bm{1}(\cdot)$ is the indicator function.
    \item \textbf{Polyphony score}, or $s^\text{poly}$, is a bit more implicit, and is defined as the average number of notes being \textit{hit} (onset) or \textit{held} (not yet released) in a sub-beat, i.e.,
    \begin{equation}\label{eq:polyphony}
        s^\text{poly} = \frac{1}{B} \sum_{b=1}^B ( n_{\text{onset}, b} + n_{\text{hold}, b} ) \,.
    \end{equation}
    This makes sense since if there are more notes pressed simultaneously, the music would feel \textit{harmonically fuller}.
\end{itemize}
After we collect all bar-wise raw scores from the dataset, we divide them into 8 bins with roughly equally many samples, resulting in the 8 classes of $a^\text{rhym}$ and $a^\text{poly}$.
For example, in our implementation, the cut-off $s^\text{rhym}$'s between the classes of $a^\text{rhym}$ are: $[.20, .25, .32, .38, .44, .50, .63]$, while the cut-offs for $a^\text{poly}$ are: $[2.63, 3.06, 3.50, 4.00, 4.63, 5.44, 6.44]$.

\textbf{Implementation and Training Details.}
Both the encoder and decoder of our MuseMorphose model comprise 12 self-attention layers \cite{vaswani2017attention}, which amount to 79.4 million trainable parameters in total.
More architectural details are displayed in Table \ref{tab:model-attr-muse-optimus} in the online supplemental materials.
For better training outcome, we introduce both \textit{cyclical KL annealing} \cite{fu2019cyclical} and \textit{free bits} \cite{kingma2016improving} (both have been introduced in Sec.~\ref{sec:vae-trick}) to our training objective (see Eq.~(\ref{eq:mopt-loss})).
After some trial and error, we set $\beta_{\text{max}} = 1$, and anneal $\beta$, i.e., the weight on the $\mathcal{L}_{\text{KL}}$ term, in cycles of 5,000 training steps.
The free bit for each dimension in $\bm{z}_k$ is set to $\lambda = 0.25\,$.
This setup is referred to as \textit{preferred settings} hereafter, for they strike a good balance between content preservation (i.e., fidelity) and diversity.\footnote{Note that this is however judged by our ears.}
We also train MuseMorphose with the pure \textit{AE objective} (i.e., $\beta=0$, constant), and \textit{VAE objective} (i.e., $\beta=1$, constant; $\lambda=0$), to examine how the model behaves on the two extremes.
\revision{Also, to see how important it is to infuse conditions into all attention layers, we include a \textit{pre-attention}\footnote{ \revision{Review Fig.~\ref{fig:transformers}a and Sec.~\ref{sec:transformers} for definition}} baseline trained under the same latent space constraints (i.e., $\beta, \lambda$) as the \textit{preferred settings}.}
Across all four settings, we do not add $\mathcal{L}_{\text{KL}}$ in the first 10,000 steps, to make it easier for the decoder to exploit the latent space in the beginning.

Furthermore, we set $K=16$ during training, meaning that we feed to the model a random 16-bar crop, truncated to maximum length $T=1{,}280$, from each musical piece in every epoch. 
This is done mainly to save memory and speed up training. 
During inference, the model can still generate music of arbitrary length with a sliding window mechanism.
As a data augmentation measure,
the key of each sample is transposed randomly in the range of $\pm6$ half notes in every epoch.

The models are trained with Adam optimizer \cite{kingma2014adam} and teacher forcing. 
We use linear warm-up to increase the learning rate to $10^{-4}$ in the first 200 steps, followed by a 200k-step cosine decay down to $5\times10^{-6}$.
Model parameters are initialized from the gaussian $\mathcal{N}(0, 0.01^2)$.
Using a batch size of 4, we can fit our MuseMorphose into a single NVIDIA Tesla V100 GPU with 32GB memory.
For all three loss variants (\textit{preferred settings}, \textit{AE objective}, \textit{VAE objective}) alike, the training converges in less than 2 full days. At inference time, we perform \textit{nucleus sampling} \cite{holtzman2019curious} to sample from the output distribution at each timestep, using a softmax temperature $\tau=1.2$ and truncating the distribution at cumulative probability $p=0.9$.

Due to the affinities in training objective and application, we implement \textit{MIDI-VAE} \cite{brunner2018midi} and \textit{Attr-Aware VAE} \cite{kawai2020attributes} as baselines.\footnote{\textit{Optimus} \cite{li2020optimus} is not included  here since it was not designed for style transfer, and that its conditioning mechanism has been evaluated in Sec.~\ref{sec:part1}.}
Their RNN-based decoders, however, operates only on single bars (i.e., $X_k$'s) of music. Therefore, during inference, each bar is generated independently and then concatenated to form the full piece.
We deem this acceptable since the bar-level latent conditions $\bm{z}_k$'s should store sufficient information to link the piece together.
We follow their specifications closely, add all auxiliary losses as required, and increase their number of trainable parameters by enlarging the RNN hidden state for fair comparison with MuseMorphose.
The resulting number of parameters in our implementations of \textit{MIDI-VAE} and \textit{Attr-Aware VAE}, respectively, are 58.2 million and 60.0 million.
For fair comparison, in our implementation the two baseline models are trained under the \textit{preferred settings} for MuseMorphose, with teacher forcing as well.

\textbf{Dataset and Data Representation.}
For this task, we consider generating expressive pop piano performances.
The pop piano MIDI dataset used is \textit{AILabs.tw-Pop1K7}, which was released in \cite{hsiao21aaai}.
According to \cite{hsiao21aaai}, the piano performances in the dataset are originally collected from the Internet
in the MP3 (audio) format. They further employed \textit{Onsets and Frames} piano transcription \cite{hawthorne2018onsets}, \texttt{madmom} beat tracking tool \cite{bock2016madmom}, and \texttt{chorder} rule-based chord detection\footnote{\url{https://github.com/joshuachang2311/chorder}} to transcribe the audio into MIDI format with tempo, beat, and chord information.
\textit{AILabs.tw-Pop1K7} encompasses 1,747 songs, or 108 hours of music, in total.
We hold out 5\% of the data (i.e., 87 songs) each for the validation and test sets.
We utilize the validation set to monitor the training process, and the test set to generate style-transferred music for further evaluation.

The data representation for songs in \textit{AILabs.tw-Pop1K7} is largely identical to REMI \cite{huang2020pop} and the one used in Sec.~\ref{sec:part1}.
Differences include: (1) an extended set of \textsc{Chord} tokens are used to mark the harmonic settings of each beat; (2) only a single piano track is present; and, (3) each bar contains 16 \textsc{Sub-beat}'s, rather than 32.
The vocabulary size here is hence reduced to 330.
Detail descriptions of all event tokens used are shown in Table~\ref{tab:event-token-remi} in the online supplemental materials.

\subsection{Evaluation Metrics}\label{sec:mmph-metrics}
To evaluate the trained models, we ask them to generate style-transferred musical pieces, i.e., $\tilde{X}$'s, based on 32-bar-long excerpts, i.e., $X$'s, drawn from the test set.\footnote{\revision{For better generation quality, during inference, we set the latent bar-level conditions $\bm{z}_k = \bm{\mu}_k$, i.e., we do not add the random noise drawn from $\bm{\sigma}_k$.}}
Following the convention in text style transfer tasks \cite{john2019disentangled, dai2019style}, the generated $\tilde{X}$'s are evaluated according to: (1) their \textit{fidelity} w.r.t.~input $X$;
(2) the strength of attribute \textit{control} given specified attributes $\tilde{a}_k^\text{rhym}$ and $\tilde{a}_k^\text{poly}$; and, (3) their \textit{fluency}.
In addition, we include a (4) \textit{diversity} criterion, which is measured across samples generated under the same inputs
$X$,  $\tilde{a}_k^\text{rhym}$ and $\tilde{a}_k^\text{poly}$.
The experiments are, therefore, conducted under two settings:
\begin{itemize}[leftmargin=*]
    \item \textbf{Setting \#1:} We randomly draw 20 excerpts from the test set, each being 32 bars long, and randomly assign to them \textit{5 sets} of different attribute inputs, i.e., $\{\tilde{a}_k^\text{rhym}, \tilde{a}_k^\text{poly}\}_{k=1}^{32}$, for a model to generate $\tilde{X}$.
    This would result in $20\cdot5=100$ samples on which we may compute the \textit{fidelity}, \textit{control}, and \textit{fluency} metrics.
    \item \textbf{Setting \#2:} We draw 20 excerpts as in setting \#1; however, here, we assign \textit{only 1 set} of attribute inputs to them each, and have the model compose \textit{5 different versions} of $\tilde{X}$ with exactly the same inputs.
    By doing so, we would obtain $20 \cdot \binom{5}{2}= 200$ pairs of generations on which we may compute the metrics on \textit{diversity}.
\end{itemize}

Concerning the metrics, for \textit{fidelity} and \textit{diversity}, we directly employ the bar-wise chroma similarity and grooving similarity (i.e., $\mathit{sim}_\text{chr}$ and $\mathit{sim}_\text{grv}$) defined in Sec.~\ref{sec_eva_obj1}. We prefer them to be higher in the case of \textit{fidelity}, but lower in the case of \textit{diversity}.
On the other hand, the metrics for \textit{control} and \textit{fluency} are defined in the following paragraphs.

\textbf{Evaluation Metrics for Attribute Control.}
Since our attributes are ordinal in nature, following \cite{kawai2020attributes}, we may directly compute the Spearman's rank correlation, $\rho$, between a specified input attribute class $a$ and the attribute raw score $s$ (see the definitions in Sec.~\ref{sec:attr-comp}, Calculation of Musical Attributes) computed from the resulting generation, to see if they are \textit{strongly} and \textit{positively} correlated.
Taking rhythmic intensity as an example, we define:
\begin{equation}
    \rho_{\text{rhym}} = \text{SpearmanCorr}(\tilde{a}^\text{rhym}, s^\text{rhym}),
\end{equation}
where $\tilde{a}^\text{rhym}$'s are user-specified inputs, and $s^\text{rhym}$'s are computed from model generations given the $\tilde{a}^\text{rhym}$'s.
The definition of $\rho_{\text{poly}}$ is similar.

Additionally, in the multi-attribute scenario, we want to avoid side effects when tuning one attribute but not the others.
More specifically, the model should be able to transfer an attribute $a$ without affecting other attributes $a'$.
To evaluate this, we define another set of correlations, $\rho_{\text{poly|rhym}}$ and $\rho_{\text{rhym|poly}}$, as:
\begin{equation}
    \rho_{\text{poly|rhym}} = \text{SpearmanCorr}(\tilde{a}^\text{rhym}, s^\text{poly}),
\end{equation}
and similarly for $\rho_{\text{rhym|poly}}$.
We prefer these correlations to be close to zero (i.e., 
$|\rho_{a'|a}|$ as small as possible), which suggest more independent control of each attribute.

\textbf{Evaluation Metric for Fluency.}
In line with text style transfer research \cite{dai2019style, john2019disentangled}, we evaluate the fluency of the generations by examining their perplexity (PPL), i.e., the exponentiation of entropy.
A low PPL means such a sequence is rather likely to be seen in a language.
(In our case, the language is pop piano music.)
However, it is impossible to know the true perplexity of a language, so our best effort is to train a language model (LM) to check instead the perplexity for that LM to generate the sequence.
This estimated perplexity 
was proven to upper-bound the true perplexity \cite{brown1992estimate}.

To this end, we train a stand-alone 24-layer Transformer decoder LM on the training set of \textit{AILabs.tw-Pop1K7}, and compute the PPL of each (32-bar-long) generation $\tilde{X}$ with:
\begin{equation}
    \text{PPL}_{\tilde{X}} = \exp( - \sum_{t=1}^{T} \log p_{\text{LM}}(\tilde{x}_t \, \vert \, \tilde{x}_{<t}) ) ,
\end{equation}
where $T$ is the length of $\tilde{X}$, and $p_{\text{LM}}(\cdot)$ is the probability given by the LM.
Note that PPL is the only piece-level metric used for our evaluation. All other metrics are measured at bar level.

\begin{table*}
\centering
\caption{Evaluation results on the style transfer of pop piano performances ($\downarrow$\,/\,$\uparrow$: the lower/higher the score the better). }\label{tab:obj-eval-vae-gen}
\begin{tabular}{=l +c+c | +c+c +c+c | +c | +c+c}
\toprule
\multirow{2}{*}{\textit{Model}}  & 
\multicolumn{2}{c|}{\textbf{Fidelity}} &
\multicolumn{4}{c|}{\textbf{Control}} &
\multicolumn{1}{c|}{\textbf{Fluency\revision{$^*$}}} &
\multicolumn{2}{c}{\textbf{Diversity}} \\
 & $\mathit{sim}_{\text{chr}}\uparrow$ & $\mathit{sim}_{\text{grv}}\uparrow$ &  $\rho_{\text{rhym}}\uparrow$ & $\rho_{\text{poly}}\uparrow$ & $|\rho_{\text{poly|rhym}}|\downarrow$ & $|\rho_{\text{rhym|poly}}|\downarrow$
& PPL$\downarrow$ & $\mathit{sim}_{\text{chr}}\downarrow$ & $\mathit{sim}_{\text{grv}}\downarrow$  \\
\midrule
\textbf{MIDI-VAE} \cite{brunner2018midi}   & 75.6 & 85.4 & .719 & .261 & .134 & .056 & 8.84 & 74.8 & 86.5  \\
\textbf{Attr-Aware VAE} \cite{kawai2020attributes} & 85.0 & 76.8 & \textbf{.997} & .781 & .239 & .040 & 10.7 & 86.5 & 84.7  \\ 
\hline
\textbf{Ours}, AE objective  & \textbf{98.5} & \textbf{95.7} & .181 & .154 & .058 & .072 & \textbf{6.10} & 97.9 & 95.4  \\ 
\textbf{Ours}, VAE objective & 78.6 & 80.7 & .931 & .884 & .038 & \textbf{.003} & 7.89 & 73.2 & 84.9  \\ 
\revtbrow \textbf{Ours}, pre-attention & 49.7 & 71.5 & .962 & \textbf{.921} & .035 & .044 & 7.33 & \textbf{45.4} & \textbf{76.8}  \\ 
\textbf{Ours}, preferred settings &  91.2 & 84.5 & .950 & .885 & \textbf{.023} & .016 & 7.39 & 87.1 & 87.6 \\ 
\bottomrule
\multicolumn{10}{l}{\revision{\footnotesize{$^*$: For reference, the fluency computed on our test set (human-composed music) is 4.79.}}} \\
\end{tabular}
\end{table*}

\begin{table}
\centering
\caption{Training and validation losses of models trained on \textit{AILabs.tw-Pop1K7} dataset. The checkpoint used is the one with the lowest validation reconstruction NLL.}\label{tab:losses-vae}
\begin{tabular}{ =l | +c+c | +c+c}
\toprule
\multirow{2}{*}{\textit{Model}}   & 
\multicolumn{2}{c|}{\textbf{Recons. NLL}} &
\multicolumn{2}{c}{\textbf{KL divergence}} \\
& \textit{Train} & \textit{Val.} & \textit{Train} & \textit{Val.} \\
\midrule
\textbf{MIDI-VAE} \cite{brunner2018midi}   & 0.676 & 0.894 & 0.697 & 0.682  \\
\textbf{Attr-Aware VAE} \cite{kawai2020attributes}  & 0.470 & 0.688 & 0.710 & 0.697   \\
\hline
\textbf{Ours}, AE objective    & 0.130 & 0.139 & 6.927 & 6.928   \\
\textbf{Ours}, VAE objective    & 0.697 & 0.765 & 0.485 & 0.484   \\
\revtbrow \textbf{Ours}, pre-attention & 0.589 & 0.860 & 0.462 & 0.474 \\
\textbf{Ours}, preferred settings   & 0.457 & 0.584 & 0.636 & 0.636   \\
\bottomrule
\end{tabular}
\end{table}

\subsection{Results and Discussion}
We commence with examining the models' losses, which are displayed in Table \ref{tab:losses-vae}.
Comparing MuseMorphose (\textit{preferred settings}) and the RNN-based baselines, which are all trained under the same latent space KL constraints (i.e., free bits and cyclical scheduling on $\beta$), 
MuseMorphose attains lower validation reconstruction NLL and KL divergence at the same time.
This is likely due to both the advantage of looking at the entire sequence, a trait that RNNs do not have; and, that a Transformer decoder can more efficiently exploit the information from latent space when paired with our \textit{in-attention} conditioning.
\revision{The latter argument can be supported by the considerably higher validation NLL of our \textit{pre-attention} baseline.}
Next, turning attention to the three \revision{in-attention variants of} MuseMorphose, we can observe clearly an inverse relationship between the amount of latent space constraint and reconstruction performance.
When the latent space is completely free (i.e., the \textit{AE objective} variant), the NLL can be reduced easily to a very low level.
Yet, even when the strict \textit{VAE objective} is used, MuseMorphose can still reach an NLL  around those of the baselines.

Table \ref{tab:obj-eval-vae-gen} shows the evaluation results on our style transfer task.
Fidelity-wise, our model mostly surpasses the baselines under the \textit{preferred settings}, except in $\mathit{sim}_{\text{grv}}$, where it scores a little lower than \textit{MIDI-VAE}.
Moreover, our \textit{AE objective} variant sticks almost perfectly to the inputs, while \textit{VAE objective} gets a much lower $\mathit{sim}_{\text{chr}}$, which is an adverse effect of strong latent space constraint. Perceptually, we find that $\mathit{sim}_{\text{chr}} < 80$ is a level where a generation often does not feel like the same piece as the input anymore.
\revision{The \textit{pre-attention} baseline, though achieving a fluency close to its in-attention counterparts, scores worse than both RNN-based methods on fidelity.
This echoes with its high reconstruction NLL, and shows that in-attention is an indispensable part of MuseMorphose.}

\begin{figure}
    \centering
    \includegraphics[width=0.42\columnwidth]{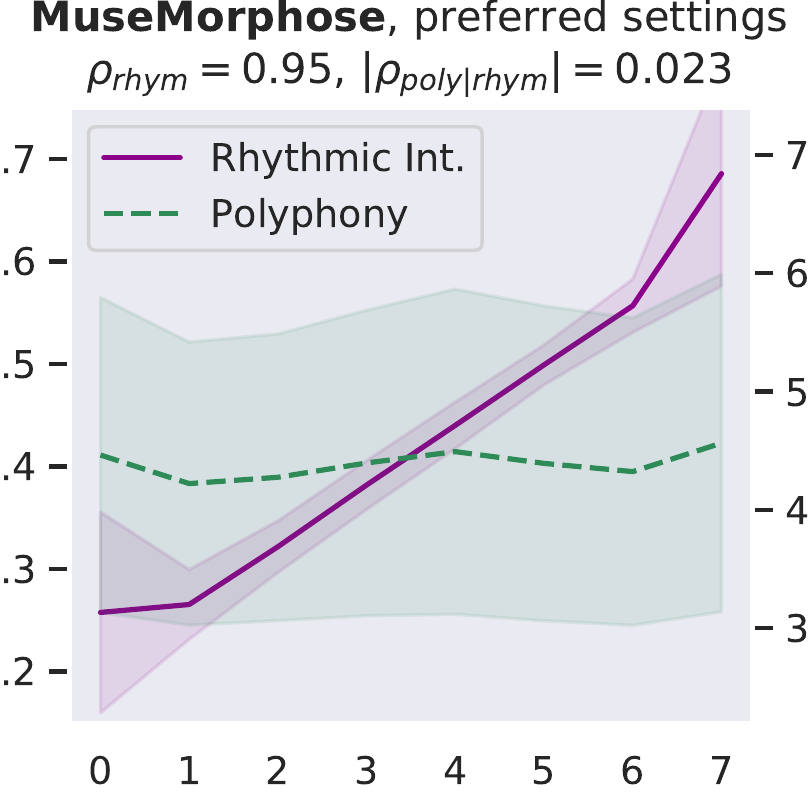}
    \hspace{3mm}
    \includegraphics[width=0.42\columnwidth]{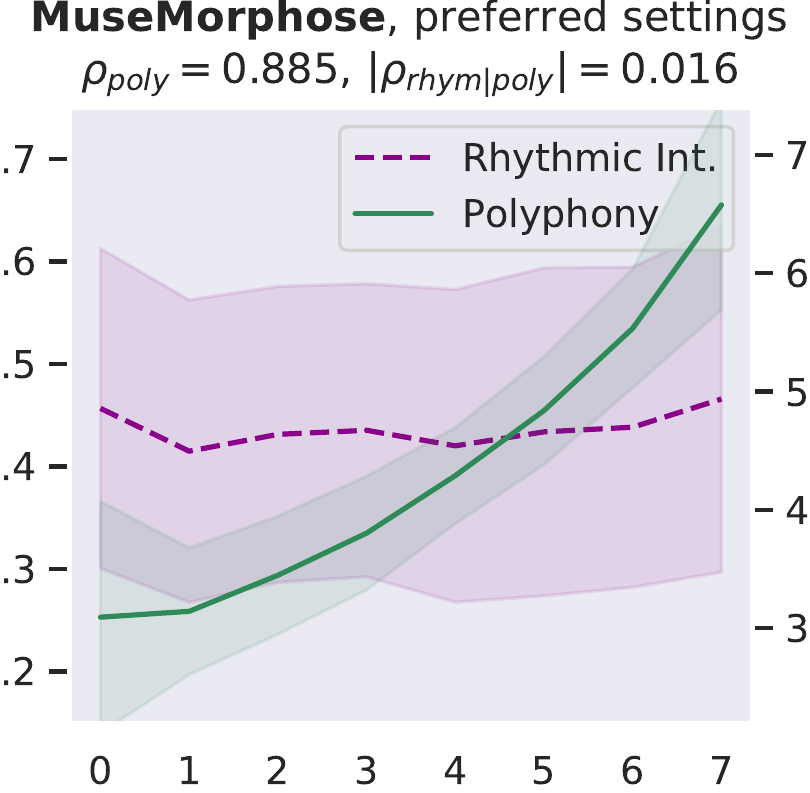}
    \caption{Mean attribute scores of MuseMorphose's generations, when controlling only rhythmic intensity (left), or only polyphony (right).
    Plots for other settings/models are in the online supplemental.
    (x-axis: controlled attribute class; y-axis left: $s^\text{rhym}$; y-axis right: $s^\text{poly}$; shaded regions represent $\pm1$ std from mean;
    \revision{class cut-offs for rhythmic intensity: $[.20, .25, .32, .38, .44, .50, .63]$; class cut-offs for polyphony: $[2.63, 3.06, 3.50, 4.00, 4.63, 5.44, 6.44]$}.)}
    \label{fig:attr-ctrl-mmph}
\end{figure}

\begin{figure*}
    \centering
    \begin{subfigure}[b]{\linewidth}
        \centering
        \includegraphics[width=0.175\textwidth]{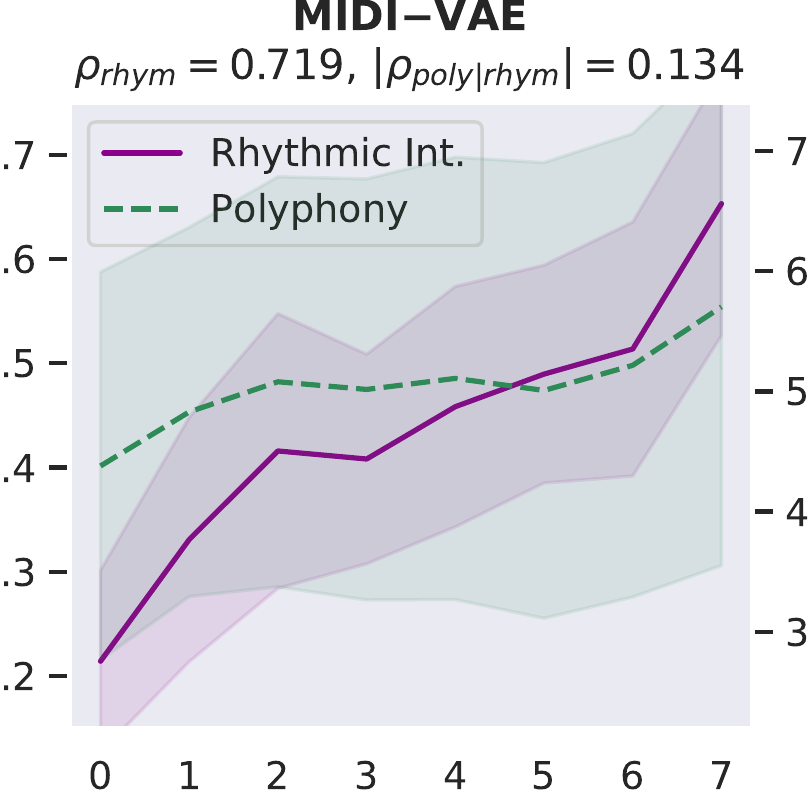}
        \hfill
        \includegraphics[width=0.175\textwidth]{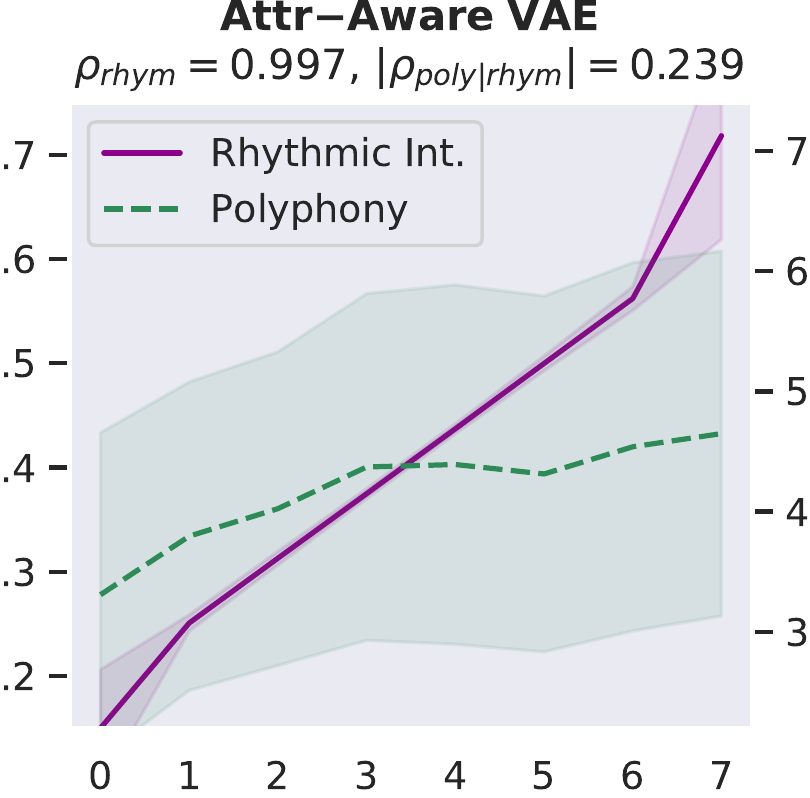}
        \hspace{1cm}
        \includegraphics[width=0.175\textwidth]{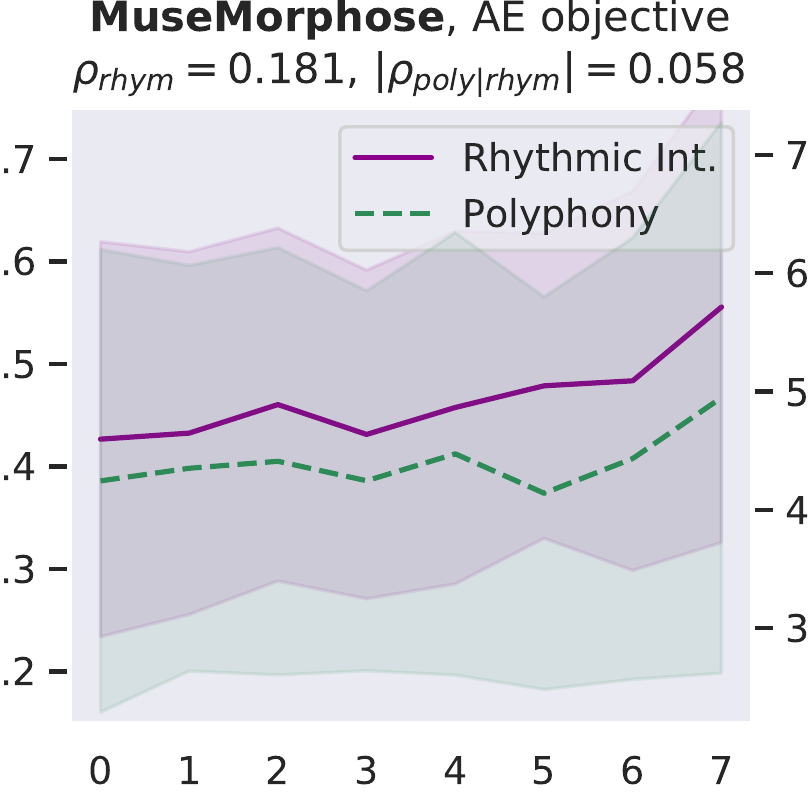}
        \hfill
        \includegraphics[width=0.175\textwidth]{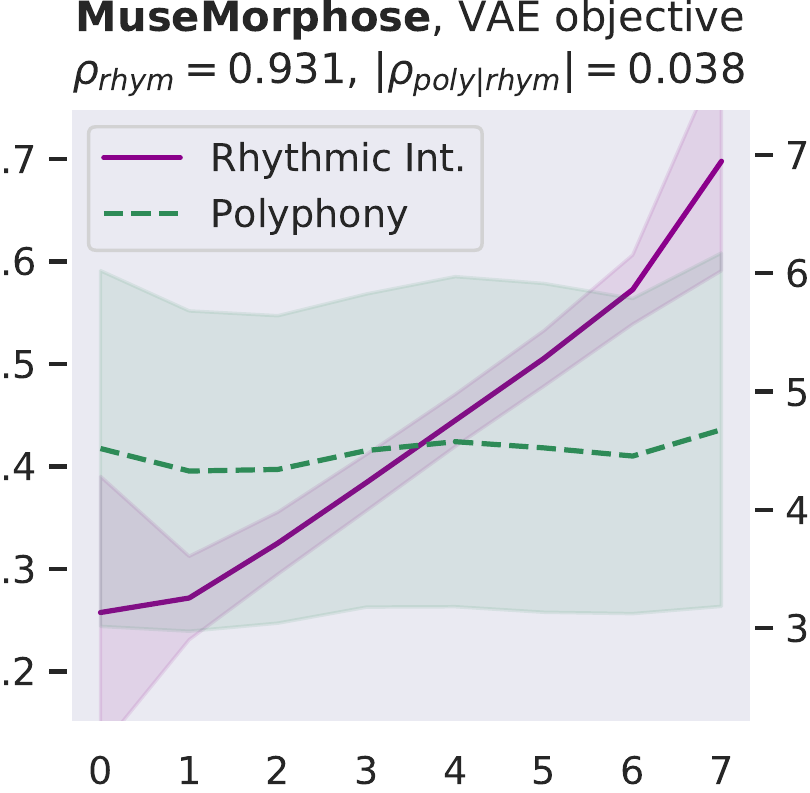}
        \hfill
        \includegraphics[width=0.175\textwidth]{figures/mmph-rhym-spearman.pdf}
        \caption{On controlling \textbf{rhythmic intensity} (\textit{x-axis}: user-specified $\tilde{a}^{\text{rhym}}$; \textit{y-axis}: $s^{\text{rhym}}$ (left), $s^{\text{poly}}$ (right) computed from the resulting generations; error bands indicate $\pm1$ std from the mean).}
        \label{subfig:rhym-ctrl}
    \end{subfigure}
    \\ \vspace{2mm}
    \centering
    \begin{subfigure}[b]{\linewidth}
        \centering
        \includegraphics[width=0.175\textwidth]{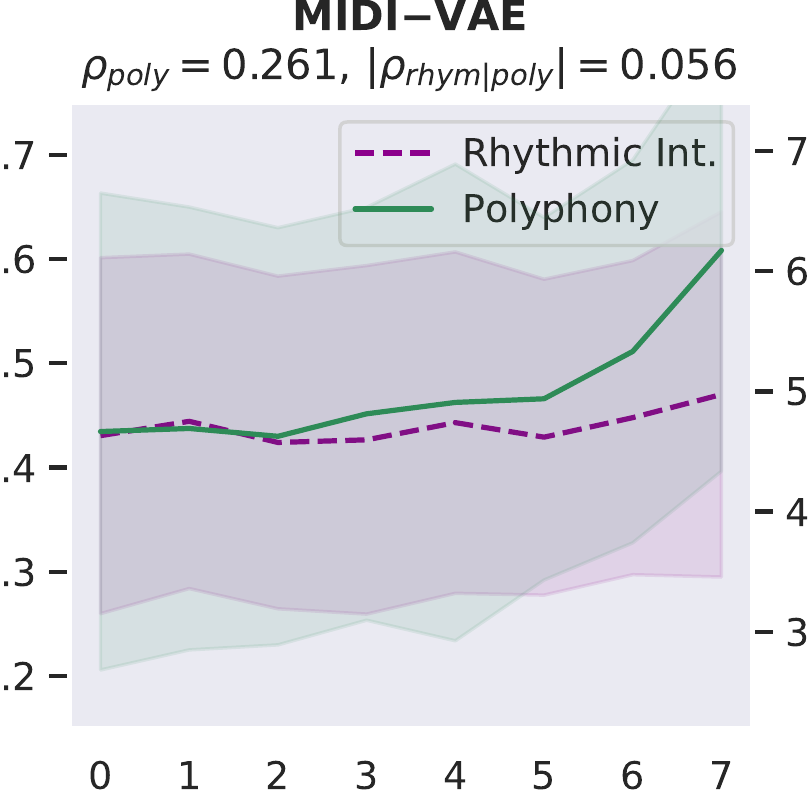}
        \hfill
        \includegraphics[width=0.175\textwidth]{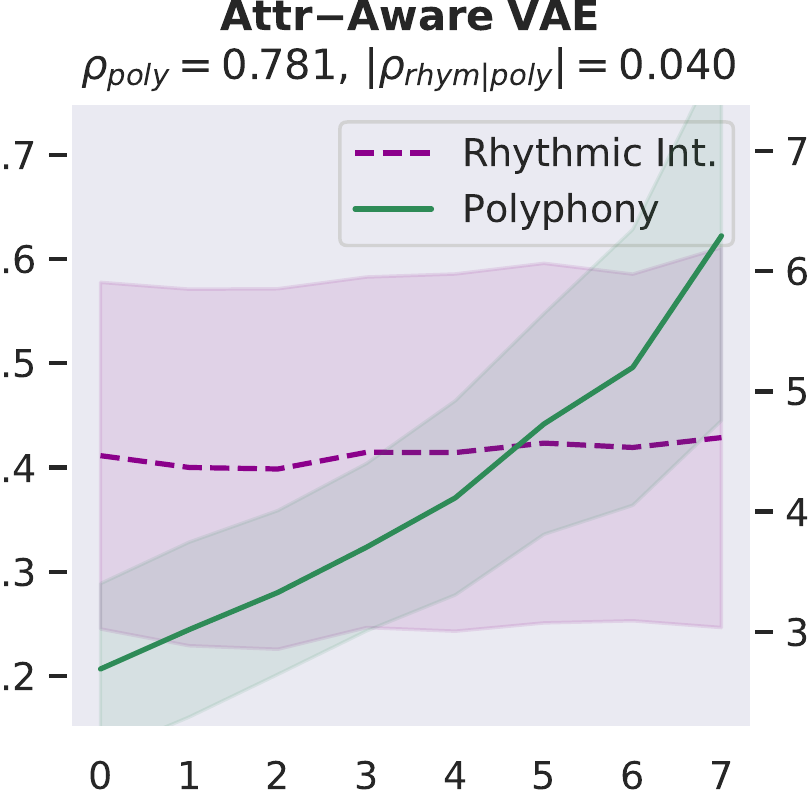}
        \hspace{1cm}
        \includegraphics[width=0.175\textwidth]{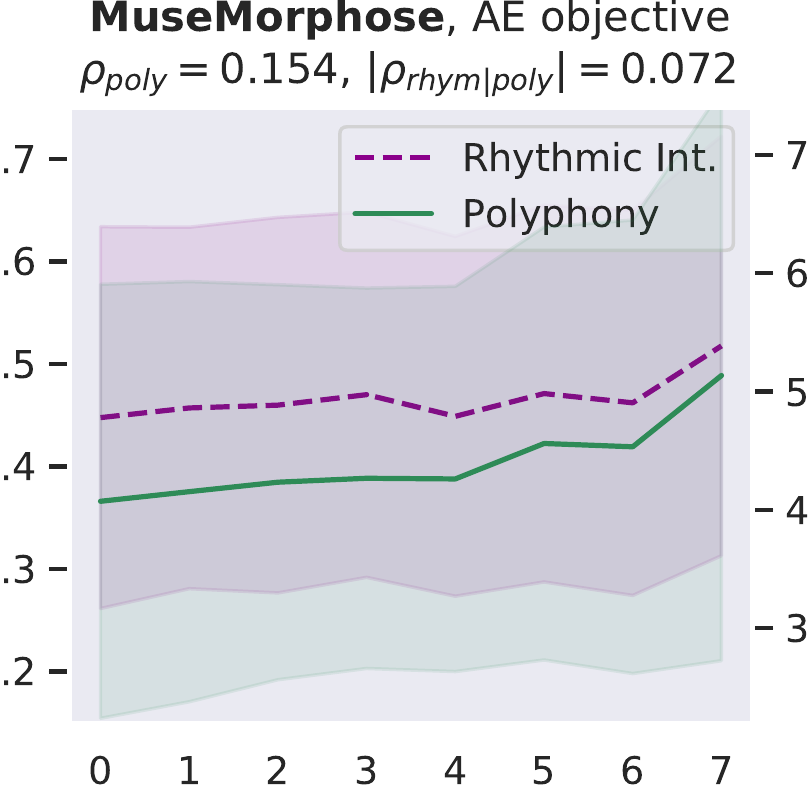}
        \hfill
        \includegraphics[width=0.175\textwidth]{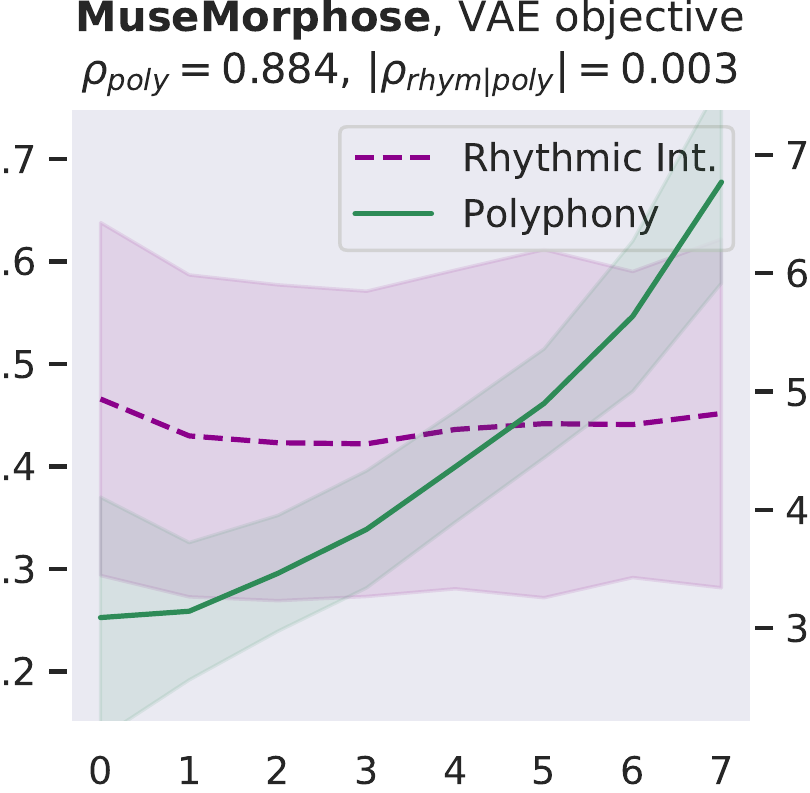}
        \hfill
        \includegraphics[width=0.175\textwidth]{figures/mmph-poly-spearman.pdf}
        \caption{On controlling \textbf{polyphony} (\textit{x-axis}: user-specified $\tilde{a}^{\text{poly}}$; \textit{y-axis}: $s^{\text{rhym}}$ (left), $s^{\text{poly}}$ (right) computed from the resulting generations; shaded regions indicate $\pm1$ std from the mean).}
        \label{subfig:poly-ctrl}
    \end{subfigure}
    \caption{Comparison of the models on attribute controllability. We desire both a high correlation $\rho_{a}$ and a low $| \rho_{a'|a} |$, where $a$ is the attribute in question, while $a'$ is not.}
    \label{fig:eval-attr-ctrl}
\end{figure*}

Attribute control is the aspect where \textit{AE objective} fails; 
meanwhile,
MuseMorphose under \textit{preferred settings} and \textit{VAE objective} exhibit equally strong attribute control with $\rho_{\text{rhym}} \approx .95$ and $\rho_{\text{poly}} \approx .90$\,.
This suggests that, in the absence of style control-targeted losses, having a narrow enough latent space bottleneck is a key to successful style control.
\revision{When the latent space is unconstrained, as in the case of \textit{AE objective}, the latent conditions may carry all necessary information for the decoder to reconstruct the music, rendering the other source of conditions, i.e., the attribute embeddings, which do not connect to the encoder, fruitless.
The KL divergence penalty that VAE imposes, on the other hand, forces disentanglement of conditions. 
It implicitly allows only the information unrelated to user-controlled attributes to stay in the latent space, thereby necessitating attribute embeddings during reconstruction.}
Comparing with the baselines, MuseMorphose is the only model capable of tightly controlling both attributes,
despite the closer competitor, \textit{Attr-Aware VAE}, achieving a really high $\rho_{\text{rhym}}$.
We surmise that rhythmic intensity is an easier attribute for the models to capture, since it only involves counting the number of \textsc{Sub-beat}'s appearing in a bar.
However, although the RNN-based methods have satisfactory control on rhythmic intensity, the polyphony scores of their generations are more or less undesirably affected ($|\rho_{\text{poly|rhym}}| \approx .15$ or $.25$) when only rhythmic intensity is tuned.
This is reasonable since the two musical attributes are correlated by nature ($|\rho_{\text{poly|rhym}}|$, and $|\rho_{\text{rhym|poly}}|$ computed from \textit{AILabs.tw-Pop1K7} are both about $.3$).
On the other hand, MuseMorphose (\textit{VAE} and \textit{preferred settings}) is able to maintain $|\rho_{\text{poly|rhym}}|$ and $|\rho_{\text{rhym|poly}}| \leq .05$ while having a firm grip on both attributes, revealing its strengths of both strong and independent attribute control.
Fig.~\ref{fig:attr-ctrl-mmph} well visualizes such strengths.
Due to space constraint, more plots comparing the attribute controllability of different settings and models are shown in Fig.~\ref{fig:eval-attr-ctrl} in the online supplemental materials.

On fluency, MuseMorphose significantly ($p < .001$)
beats the previous works under all settings.
Diversity-wise, though our model does not seem to have an edge over them, we note that there exists an inherent trade-off between diversity and fidelity.
If we compare our \textit{preferred settings} with Attr-Aware VAE, the closer competitor, we can notice that the margins by which MuseMorphose wins over Attr-Aware VAE on fidelity ($6.2$ for $\mathit{sim}_{\text{chr}}$; $7.7$ for  $\mathit{sim}_{\text{grv}}$) is somewhat larger than those by which it loses on diversity ($0.6$ for $\mathit{sim}_{\text{chr}}$; $2.9$ for  $\mathit{sim}_{\text{grv}}$).

\revision{It is worth mentioning that manipulating the fidelity-diversity is possible both at training and inference time.
During training, we may adjust VAE-related hyperparameters like $\beta$ and free bits ($\lambda$). The comparison between \textit{preferred settings} and \textit{VAE objective} offers us a glimpse of the resulting effect.
During inference, to boost diversity, we may draw noises from std vectors (i.e., $\bm{\sigma}_k$'s, which may be enlarged or shrunk by multiplying by a constant) and add them to latent conditions, increase sampling temperature, or, alter the distribution truncation point for nucleus sampling. The respective impact of these measures is an interesting direction for future studies.
}

\begin{figure}
    \centering
    \includegraphics[width=0.48\linewidth]{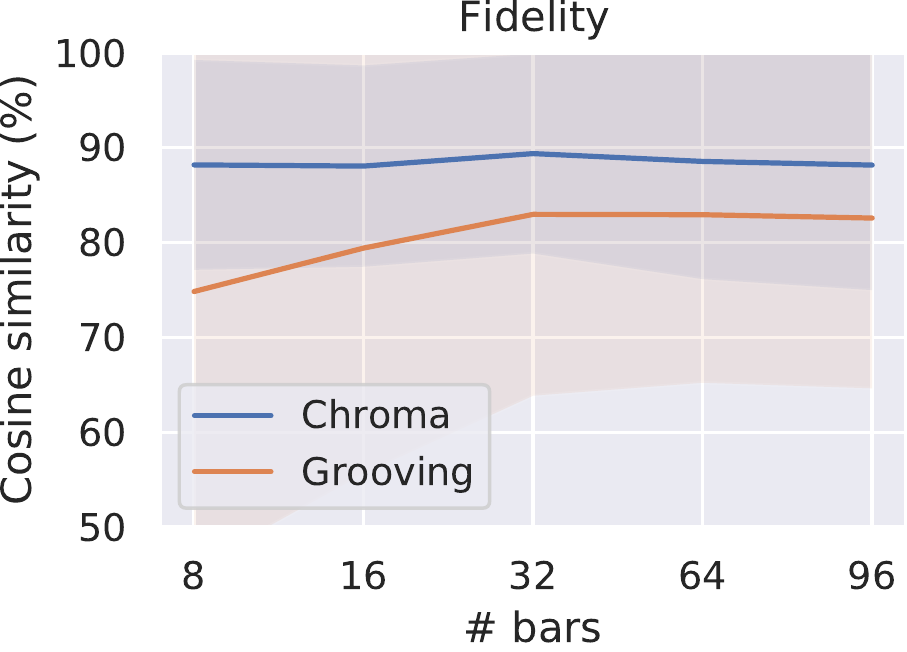}
    \hfill
    \includegraphics[width=0.48\linewidth]{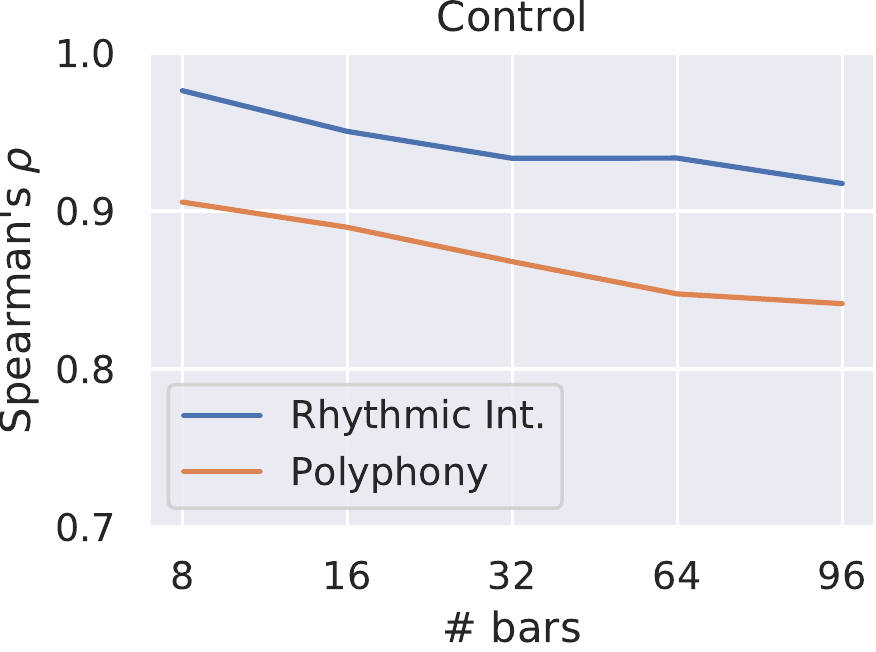}
    \\
    \vspace{5mm}
    \includegraphics[width=0.48\linewidth]{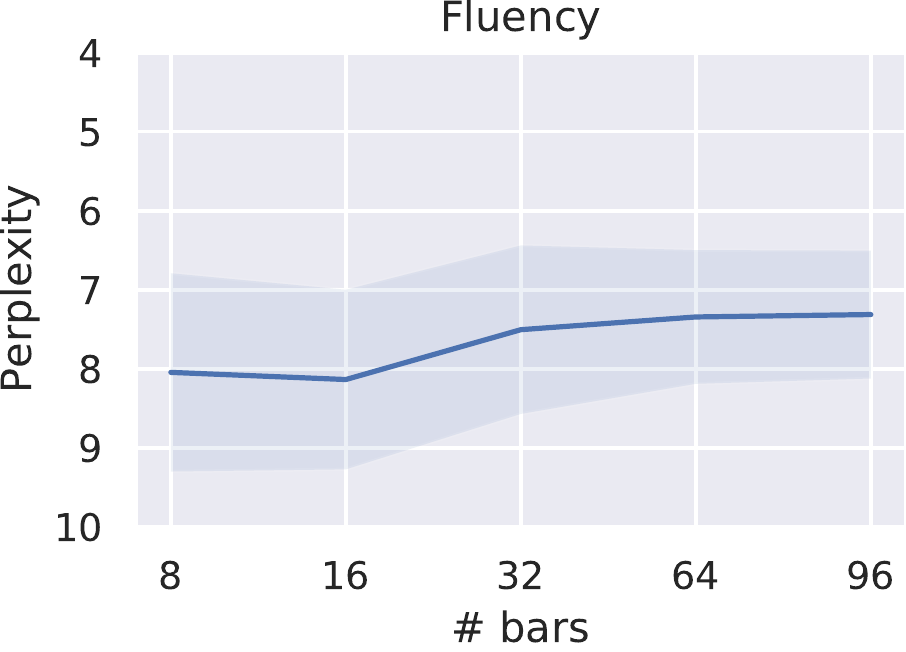}
    \hfill
    \includegraphics[width=0.48\linewidth]{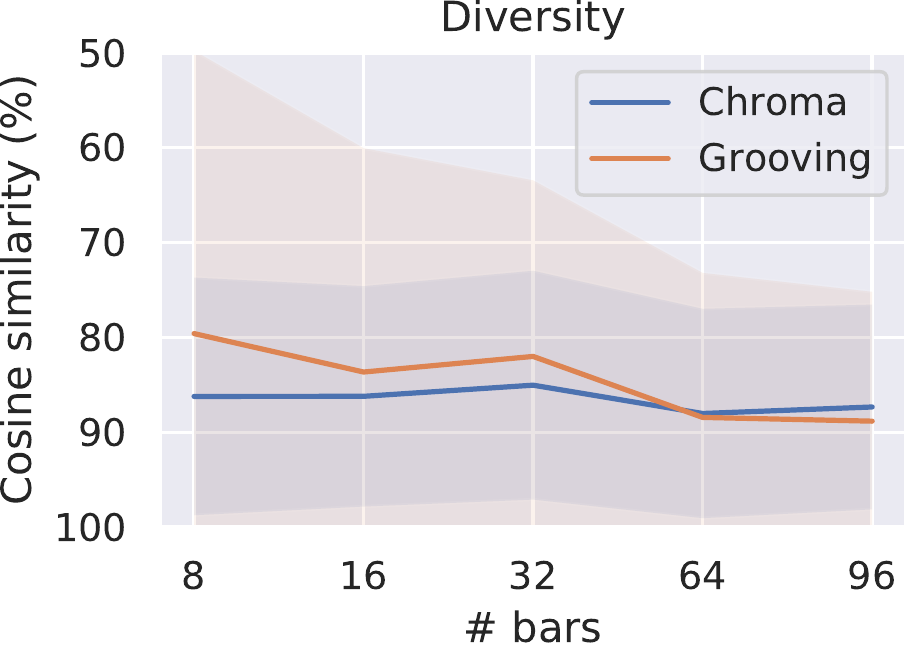}
    \caption{Evaluation on MuseMorphose's generations of different lengths. (Y-axis for ``Fluency'' and ``Diversity'' plots are inverted since these two metrics are the lower the better; shaded regions indicate $\pm1$ std from the mean. Best viewed in color.)}
    \label{fig:eval-var-length}
\end{figure}

\revision{To show that MuseMorphose is able to perform style transfer on arbitrarily long music excerpts, we run evaluation on 8$\sim$96 bar-long generations by our \textit{preferred settings} model. The results are plotted in Fig.~\ref{fig:eval-var-length} in the online supplemental. The scores stay rather consistent as generation length increases, except for slight downward trends of attribute control and grooving diversity. These trends are reasonable since the higher number of bar-level blueprints given to the model could collectively place stronger restrictions on the generated content.}



To summarize, our MuseMorphose model, underpinned by Transformers and the in-attention conditioning mechanism, outperforms both baselines and ticks all the boxes in our controllable music generation task---in particular, it accomplishes high \textit{fidelity}, strong and independent \textit{attribute control}, good sequence-level \textit{fluency}, and adequate \textit{diversity}, all at once.
Nevertheless, we also emphasize that using a combination of VAE training techniques, i.e., cyclical KL annealing and free bits, and picking suitable values for the hyperparameters $\beta$ and $\lambda$, are indispensable to the success of MuseMorphose.

\section{Conclusion and Future Directions}
In this paper, we developed \textit{MuseMorphose}, a sequence variational autoencoder with Transformers as its backbone, which delivered a stellar performance on controllable conditional generation of pop piano performances.
In achieving so, we set off from defining a novel segment-level conditioning problem for generative Transformers, and devised three mechanisms, namely, \textit{pre-attention}, \textit{in-attention}, and \textit{post-attention} conditioning, to approach it.
Conducted objective evaluations demonstrated that in-attention came out on top in terms of offering firm control with time-varying conditioning vectors.
Subsequently, we leveraged in-attention to bring together a Transformer encoder and Transformer decoder, forming the foundation of MuseMorphose model.
Experiments have shown that, when trained with a carefully tuned VAE objective, MuseMorphose emerged as a solid all-rounder on the music style transfer task with long inputs,
where we considered controlling two musical attributes, i.e., \textit{rhythmic intensity} and \textit{polyphony}, at the bar level.
Our model outperformed two previous methods \cite{brunner2018midi, kawai2020attributes} on commonly accepted style transfer evaluation metrics, without using any auxiliary objectives tailored for style control.

\revrdtwo{Nevertheless, whether in-attention would perform best across more kinds of musical conditions (e.g., tonality, instrumentation, which vary less frequently than the two studied attributes), and why it achieves so remain underexplored in the current work.
Besides, it is somewhat unclear how much performance gain actually comes from replacing the RNN sequence models with Transformers, rather than from our proposed conditioning and training mechanisms.
Experimenting on shorter music excerpts (e.g., 2$\sim$4 bars, which RNNs can process at once) may shed light on this.
We hope future studies may address the unanswered aspects above.}

Our research can also be extended in the following directions:
\begin{itemize}[leftmargin=*]
    \item \textbf{Applications in Natural Language Processing (NLP):} The framework of MuseMorphose is easily generalizable to texts by, for example, treating \textit{sentences} as segments (i.e., $X_k$'s) and an \textit{article} as the full sequence (i.e., $X$). As for attributes, possible options include \textit{sentiments} \cite{fu2018style, dai2019style} and \textit{text difficulty} \cite{surya2019unsupervised}.
    Success in controlling the latter, for example, is of high practical value since it would enable us to rewrite articles or literary works to cater to the needs of pupils across all stages of education.
    \item \textbf{Music generation with long-range structures:} 
    Being inspired by the two astounding works of high-resolution image generation \cite{ramesh2021zero, esser2021taming}, in which Transformers are used to generate the high-level latent semantics of an image (i.e., $\bm{z}_k$'s, the sequence of latent conditions),
    we conjecture that long range musical structures, e.g., motivic development, recurrence of themes, contrasting sections, etc., may be better generated in a similar fashion.
    To this end, it is likely that we need to learn a \textit{vector-quantized} latent space \cite{van2017neural} instead, so that the latent conditions can be represented in token form for another Transformer to model.
\end{itemize}

To conclude, our work not only bridged Transformers and VAEs for controllable sequence modeling, where potentials for further applications exist, but also laid the 
foundation for a pathway to long sequence generation never explored before.
\ifCLASSOPTIONcaptionsoff
  \newpage
\fi



\bibliographystyle{IEEEtran}
\bibliography{references.bib}

\begin{thebibliography}{10}
\providecommand{\url}[1]{#1}
\csname url@samestyle\endcsname
\providecommand{\newblock}{\relax}
\providecommand{\bibinfo}[2]{#2}
\providecommand{\BIBentrySTDinterwordspacing}{\spaceskip=0pt\relax}
\providecommand{\BIBentryALTinterwordstretchfactor}{4}
\providecommand{\BIBentryALTinterwordspacing}{\spaceskip=\fontdimen2\font plus
\BIBentryALTinterwordstretchfactor\fontdimen3\font minus
  \fontdimen4\font\relax}
\providecommand{\BIBforeignlanguage}[2]{{%
\expandafter\ifx\csname l@#1\endcsname\relax
\typeout{** WARNING: IEEEtran.bst: No hyphenation pattern has been}%
\typeout{** loaded for the language `#1'. Using the pattern for}%
\typeout{** the default language instead.}%
\else
\language=\csname l@#1\endcsname
\fi
#2}}
\providecommand{\BIBdecl}{\relax}
\BIBdecl

\bibitem{hiller1959experiment}
L.~A. Hiller and L.~M. Isaacson, \emph{Experimental Music; Composition with an
  Electronic Computer}, 1959.

\bibitem{hadjeres2017deepbach}
G.~Hadjeres, F.~Pachet, and F.~Nielsen, ``Deep{B}ach: a steerable model for
  {B}ach chorales generation,'' in \emph{Proc. ICML}, 2017.

\bibitem{dong2018musegan}
H.-W. Dong, W.-Y. Hsiao, L.-C. Yang, and Y.-H. Yang, ``Muse{GAN}: Multi-track
  sequential generative adversarial networks for symbolic music generation and
  accompaniment,'' in \emph{Proc. AAAI}, 2018.

\bibitem{ji20survey}
S.~Ji, J.~Luo, and X.~Yang, ``A comprehensive survey on deep music generation:
  Multi-level representations, algorithms, evaluations, and future
  directions,'' \emph{arXiv preprint arXiv:2011.06801}, 2020.

\bibitem{hawthorne2018onsets}
C.~Hawthorne, E.~Elsen, J.~Song, A.~Roberts, I.~Simon, C.~Raffel, J.~Engel,
  S.~Oore, and D.~Eck, ``{Onsets and Frames}: Dual-objective piano
  transcription,'' in \emph{Proc. ISMIR}, 2018.

\bibitem{oore2018time}
S.~Oore, I.~Simon, S.~Dieleman, D.~Eck, and K.~Simonyan, ``This time with
  feeling: Learning expressive musical performance,'' \emph{arXiv preprint
  arXiv:1808.03715}, 2018.

\bibitem{huang2020pop}
Y.-S. Huang and Y.-H. Yang, ``{Pop Music Transformer}: {Generating} music with
  rhythm and harmony,'' in \emph{Proc. ACM Multimedia}, 2020.

\bibitem{sennrich2016neural}
R.~Sennrich, B.~Haddow, and A.~Birch, ``Neural machine translation of rare
  words with subword units,'' in \emph{Proc. ACL}, 2016.

\bibitem{huang19music}
C.~A. Huang, A.~Vaswani, J.~Uszkoreit, I.~Simon, C.~Hawthorne, N.~Shazeer,
  A.~M. Dai, M.~D. Hoffman, M.~Dinculescu, and D.~Eck, ``Music {T}ransformer:
  Generating music with long-term structure,'' in \emph{Proc. {ICLR}}, 2019.

\bibitem{donahue2019lakhnes}
C.~Donahue, H.~H. Mao, Y.~E. Li, G.~W. Cottrell, and J.~McAuley, ``Lakh{NES}:
  Improving multi-instrumental music generation with cross-domain
  pre-training,'' in \emph{Proc. ISMIR}, 2019.

\bibitem{roberts2018hierarchical}
A.~Roberts, J.~Engel, C.~Raffel, C.~Hawthorne, and D.~Eck, ``A hierarchical
  latent vector model for learning long-term structure in music,'' in
  \emph{Proc. ICML}, 2018.

\bibitem{vaswani2017attention}
A.~Vaswani, N.~Shazeer, N.~Parmar, J.~Uszkoreit, L.~Jones, A.~N. Gomez,
  L.~Kaiser, and I.~Polosukhin, ``Attention is all you need,'' in \emph{Proc.
  NeurIPS}, 2017.

\bibitem{kingma2014auto}
D.~P. Kingma and M.~Welling, ``Auto-encoding variational bayes,'' in
  \emph{Proc. ICLR}, 2014.

\bibitem{hochreiter97lstm}
S.~Hochreiter and J.~Schmidhuber, ``Long short-term memory,'' \emph{Neural
  Computation}, vol.~9, no.~8, pp. 1735--1780, 1997.

\bibitem{cho-etal-2014-properties}
K.~Cho, B.~van Merri{\"e}nboer, D.~Bahdanau, and Y.~Bengio, ``On the properties
  of neural machine translation: Encoder{--}decoder approaches,'' in
  \emph{Proc. Eighth Workshop on Syntax, Semantics and Structure in Statistical
  Translation}, 2014.

\bibitem{payne2019musenet}
C.~M. Payne, ``{MuseNet},'' \emph{OpenAI Blog}, 2019.

\bibitem{wu2020jazz}
S.-L. Wu and Y.-H. Yang, ``The {J}azz {T}ransformer on the front line:
  Exploring the shortcomings of {AI}-composed music through quantitative
  measures,'' in \emph{Proc. ISMIR}, 2020.

\bibitem{chen2020automatic}
Y.-H. Chen, Y.-H. Huang, W.-Y. Hsiao, and Y.-H. Yang, ``Automatic composition
  of guitar tabs by {T}ransformers and groove modeling,'' in \emph{Proc.
  ISMIR}, 2020.

\bibitem{dai2019transformer}
Z.~Dai, Z.~Yang, Y.~Yang, J.~G. Carbonell, Q.~Le, and R.~Salakhutdinov,
  ``Transformer-{XL}: Attentive language models beyond a fixed-length
  context,'' in \emph{Proc. ACL}, 2019.

\bibitem{tan2020music}
H.~H. Tan and D.~Herremans, ``Music {F}ader{N}ets: Controllable music
  generation based on high-level features via low-level feature modelling,'' in
  \emph{Proc. ISMIR}, 2020.

\bibitem{hadjeres2017glsr}
G.~Hadjeres, F.~Nielsen, and F.~Pachet, ``{GLSR-VAE}: Geodesic latent space
  regularization for variational autoencoder architectures,'' in \emph{Proc.
  IEEE Symposium Series on Computational Intelligence}, 2017.

\bibitem{brunner2018midi}
G.~Brunner, A.~Konrad, Y.~Wang, and R.~Wattenhofer, ``{MIDI-VAE}: Modeling
  dynamics and instrumentation of music with applications to style transfer,''
  in \emph{Proc. ISMIR}, 2018.

\bibitem{kawai2020attributes}
L.~Kawai, P.~Esling, and T.~Harada, ``Attributes-aware deep music
  transformation,'' in \emph{Proc. ISMIR}, 2020.

\bibitem{goodfellow2014generative}
I.~J. Goodfellow, J.~Pouget-Abadie, M.~Mirza, B.~Xu, D.~Warde-Farley, S.~Ozair,
  A.~Courville, and Y.~Bengio, ``Generative adversarial networks,'' in
  \emph{Proc. NeurIPS}, 2014.

\bibitem{choi2020encoding}
K.~Choi, C.~Hawthorne, I.~Simon, M.~Dinculescu, and J.~Engel, ``Encoding
  musical style with {T}ransformer autoencoders,'' in \emph{Proc. ICML}, 2020.

\bibitem{ren2020popmag}
Y.~Ren, J.~He, X.~Tan, T.~Qin, Z.~Zhao, and T.-Y. Liu, ``Pop{MAG}: Pop music
  accompaniment generation,'' in \emph{Proc. ACM Multimedia}, 2020.

\bibitem{hsiao21aaai}
W.-Y. Hsiao, J.-Y. Liu, Y.-C. Yeh, and Y.-H. Yang, ``{Compound Word
  Transformer}: {Learning} to compose full-song music over dynamic directed
  hypergraphs,'' in \emph{Proc. AAAI}, 2021.

\bibitem{radford2019language}
A.~Radford, J.~Wu, R.~Child, D.~Luan, D.~Amodei, and I.~Sutskever, ``Language
  models are unsupervised multitask learners,'' \emph{OpenAI Blog}, 2019.

\bibitem{devlin18bert}
J.~Devlin, M.-W. Chang, K.~Lee, and K.~N. Toutanova, ``{BERT}: Pre-training of
  deep bidirectional {T}ransformers for language understanding,'' in
  \emph{Proc. NAACL-HLT}, 2018.

\bibitem{fu2018style}
Z.~Fu, X.~Tan, N.~Peng, D.~Zhao, and R.~Yan, ``Style transfer in text:
  Exploration and evaluation,'' in \emph{Proc. AAAI}, 2018.

\bibitem{panda2018novel}
R.~Panda, R.~Malheiro, and R.~P. Paiva, ``Novel audio features for music
  emotion recognition,'' \emph{IEEE Transactions on Affective Computing}, 2018.

\bibitem{panda2020audio}
R.~Panda, R.~M. Malheiro, and R.~P. Paiva, ``Audio features for music emotion
  recognition: a survey,'' \emph{IEEE Transactions on Affective Computing},
  2020.

\bibitem{li2020optimus}
C.~Li, X.~Gao, Y.~Li, B.~Peng, X.~Li, Y.~Zhang, and J.~Gao, ``Optimus:
  Organizing sentences via pre-trained modeling of a latent space,'' in
  \emph{Proc. EMNLP}, 2020.

\bibitem{wang2019t}
T.~Wang and X.~Wan, ``{T-CVAE}: Transformer-based conditioned variational
  autoencoder for story completion.'' in \emph{Proc. IJCAI}, 2019.

\bibitem{he2016deep}
K.~He, X.~Zhang, S.~Ren, and J.~Sun, ``Deep residual learning for image
  recognition,'' in \emph{Proc. CVPR}, 2016.

\bibitem{ba2016layer}
J.~L. Ba, J.~R. Kiros, and G.~E. Hinton, ``Layer normalization,'' \emph{arXiv
  preprint arXiv:1607.06450}, 2016.

\bibitem{shaw2018self}
P.~Shaw, J.~Uszkoreit, and A.~Vaswani, ``Self-attention with relative position
  representations,'' in \emph{Proc. NAACL}, 2018.

\bibitem{ke2021rethinking}
G.~Ke, D.~He, and T.-Y. Liu, ``Rethinking the positional encoding in language
  pre-training,'' in \emph{Proc. ICLR}, 2021.

\bibitem{wang2021position}
B.~Wang, L.~Shang, C.~Lioma, X.~Jiang, H.~Yang, Q.~Liu, and J.~G. Simonsen,
  ``On position embeddings in bert,'' in \emph{Proc. ICLR}, 2021.

\bibitem{liutkus2021relative}
A.~Liutkus, O.~C{\'i}fka, S.-L. Wu, U.~{\c S}im{\c s}ekli, Y.-H. Yang, and
  G.~Richard, ``Relative positional encoding for {T}ransformers with linear
  complexity,'' in \emph{Proc. ICML}, 2021.

\bibitem{katharopoulos2020transformers}
A.~Katharopoulos, A.~Vyas, N.~Pappas, and F.~Fleuret, ``Transformers are
  {RNN}s: Fast autoregressive {T}ransformers with linear attention,'' in
  \emph{Proc. ICML}, 2020.

\bibitem{choromanski2021rethinking}
K.~Choromanski, V.~Likhosherstov, D.~Dohan, X.~Song, A.~Gane, T.~Sarlos,
  P.~Hawkins, J.~Davis, A.~Mohiuddin, L.~Kaiser \emph{et~al.}, ``Rethinking
  attention with performers,'' in \emph{Proc. ICLR}, 2021.

\bibitem{higgins2017beta}
I.~Higgins, L.~Matthey, A.~Pal, C.~Burgess, X.~Glorot, M.~Botvinick,
  S.~Mohamed, and A.~Lerchner, ``beta-{VAE}: Learning basic visual concepts
  with a constrained variational framework,'' in \emph{Proc. ICLR}, 2017.

\bibitem{sonderby2016ladder}
C.~K. S{\o}nderby, T.~Raiko, L.~Maal{\o}e, S.~K. S{\o}nderby, and O.~Winther,
  ``Ladder variational autoencoders,'' in \emph{Proc. NeurIPS}, 2016.

\bibitem{bowman2016generating}
S.~Bowman, L.~Vilnis, O.~Vinyals, A.~Dai, R.~Jozefowicz, and S.~Bengio,
  ``Generating sentences from a continuous space,'' in \emph{Proc. SIGNLL},
  2016.

\bibitem{dieng2019avoiding}
A.~B. Dieng, Y.~Kim, A.~M. Rush, and D.~M. Blei, ``Avoiding latent variable
  collapse with generative skip models,'' in \emph{Proc. AISTATS}, 2019.

\bibitem{kingma2016improving}
D.~P. Kingma, T.~Salimans, R.~Jozefowicz, X.~Chen, I.~Sutskever, and
  M.~Welling, ``Improving variational inference with inverse autoregressive
  flow,'' \emph{arXiv preprint arXiv:1606.04934}, 2016.

\bibitem{fu2019cyclical}
H.~Fu, C.~Li, X.~Liu, J.~Gao, A.~Celikyilmaz, and L.~Carin, ``Cyclical
  annealing schedule: A simple approach to mitigating {KL} vanishing,'' in
  \emph{Proc. NAACL-HLT}, 2019.

\bibitem{keskar2019ctrl}
N.~S. Keskar, B.~McCann, L.~R. Varshney, C.~Xiong, and R.~Socher, ``{CTRL}: A
  conditional {T}ransformer language model for controllable generation,''
  \emph{arXiv preprint arXiv:1909.05858}, 2019.

\bibitem{stern2019insertion}
M.~Stern, W.~Chan, J.~Kiros, and J.~Uszkoreit, ``{Insertion Transformer}:
  Flexible sequence generation via insertion operations,'' in \emph{Proc.
  ICML}, 2019.

\bibitem{kingma2014adam}
D.~P. Kingma and J.~Ba, ``Adam: A method for stochastic optimization,''
  \emph{arXiv preprint arXiv:1412.6980}, 2014.

\bibitem{chen2020generative}
M.~Chen, A.~Radford, J.~Wu, H.~Jun, P.~Dhariwal, D.~Luan, and I.~Sutskever,
  ``Generative pretraining from pixels,'' in \emph{Proc. ICML}, 2020.

\bibitem{bmusegan}
H.-W. Dong and Y.-H. Yang, ``Convolutional generative adversarial networks with
  binary neurons for polyphonic music generation,'' in \emph{Proc. ISMIR},
  2018.

\bibitem{fujishima99}
T.~Fujishima, ``Realtime chord recognition of musical sound: A system using
  common {L}isp,'' in \emph{Proc. International Computer Music Conf. (ICMC)},
  1999.

\bibitem{dixonEtAl04ismir}
S.~Dixon, F.~Gouyon, and G.~Widmer, ``Towards characterisation of music via
  rhythmic patterns,'' in \emph{Proc. ISMIR}, 2004.

\bibitem{foote1999visualizing}
J.~Foote, ``Visualizing music and audio using self-similarity,'' in \emph{Proc.
  ACM Multimedia}, 1999.

\bibitem{yang20evaluation}
L.-C. Yang and A.~Lerch, ``On the evaluation of generative models in music,''
  \emph{Neural Computing and Applications}, vol.~32, no.~9, pp. 4773--4784,
  2020.

\bibitem{yang2018unsupervised}
Z.~Yang, Z.~Hu, C.~Dyer, E.~P. Xing, and T.~Berg-Kirkpatrick, ``Unsupervised
  text style transfer using language models as discriminators,'' in \emph{Proc.
  NeurIPS}, 2018.

\bibitem{john2019disentangled}
V.~John, L.~Mou, H.~Bahuleyan, and O.~Vechtomova, ``Disentangled representation
  learning for non-parallel text style transfer,'' in \emph{Proc. ACL}, 2019.

\bibitem{lample2018multiple}
G.~Lample, S.~Subramanian, E.~Smith, L.~Denoyer, M.~Ranzato, and Y.-L. Boureau,
  ``Multiple-attribute text rewriting,'' in \emph{Proc. ICLR}, 2019.

\bibitem{dai2019style}
N.~Dai, J.~Liang, X.~Qiu, and X.-J. Huang, ``Style {T}ransformer: Unpaired text
  style transfer without disentangled latent representation,'' in \emph{Proc.
  ACL}, 2019.

\bibitem{holtzman2019curious}
A.~Holtzman, J.~Buys, L.~Du, M.~Forbes, and Y.~Choi, ``The curious case of
  neural text degeneration,'' in \emph{Proc. ICLR}, 2019.

\bibitem{bock2016madmom}
S.~B{\"o}ck, F.~Korzeniowski, J.~Schl{\"u}ter, F.~Krebs, and G.~Widmer,
  ``Madmom: {A} new {Python} audio and music signal processing library,'' in
  \emph{Proc. ACM Multimedia}, 2016.

\bibitem{brown1992estimate}
P.~F. Brown, S.~A. Della~Pietra, V.~J. Della~Pietra, J.~C. Lai, and R.~L.
  Mercer, ``An estimate of an upper bound for the entropy of english,''
  \emph{Computational Linguistics}, vol.~18, no.~1, pp. 31--40, 1992.

\bibitem{surya2019unsupervised}
S.~Surya, A.~Mishra, A.~Laha, P.~Jain, and K.~Sankaranarayanan, ``Unsupervised
  neural text simplification,'' in \emph{Proc. ACL}, 2019.

\bibitem{ramesh2021zero}
A.~Ramesh, M.~Pavlov, G.~Goh, S.~Gray, C.~Voss, A.~Radford, M.~Chen, and
  I.~Sutskever, ``Zero-shot text-to-image generation,'' \emph{arXiv preprint
  arXiv:2102.12092}, 2021.

\bibitem{esser2021taming}
P.~Esser, R.~Rombach, and B.~Ommer, ``Taming {T}ransformers for high-resolution
  image synthesis,'' in \emph{Proc. CVPR}, 2021.

\bibitem{van2017neural}
A.~van~den Oord, O.~Vinyals, and K.~Kavukcuoglu, ``Neural discrete
  representation learning,'' in \emph{Proc. NeurIPS}, 2017.

\end{thebibliography}
%

\clearpage


\section{Supplemental Materials:\\Extra Tables and Figures}
\begin{table}[h]
\centering
\caption{Main attributes shared by the implemented Transformer decoders in Section \ref{sec:part1}.}\label{tab:model-attr}
\begin{tabular}{l l r}
\toprule
Attr. & Description & Value \\
\midrule
$T_{\text{tgt}}$ & target sequence length & 1,024 \\
$T_{\text{mem}}$ & Transformer-XL memory length & 1,024 \\
$L$ & \# self-attention layers & 12 \\
$n_{\text{head}}$ & \# self-attention heads & 10 \\
$d_{\text{e}}$ & token embedding dimension & 320 \\
$d$ & hidden state dimension & 640 \\
$d_{\text{ff}}$ & feed-forward dimension & 2,048 \\
$d_{\text{c}}$ & condition embedding dimension & 512 \\
\hline
\# params & \multicolumn{2}{r}{58.7$\sim$62.6 mil.}\\
\bottomrule
\end{tabular}
\end{table}

\begin{table}[h]
\centering
\caption{Main attributes of our \textit{MuseMorphose} model.}\label{tab:model-attr-muse-optimus}
\begin{tabular}{l l r}
\toprule
Attr. & Description & Value \\
\midrule
$T$ & target sequence length & 1,280 \\
$L$ & \# self-attention layers & 24 \\
$L_\textbf{enc}$ & \# encoder self-attention layers & 12 \\
$L_\textbf{dec}$ & \# decoder self-attention layers & 12 \\
$n_{\text{head}}$ & \# self-attention heads & 8 \\
$d_{\text{e}}$ & token embedding dimension & 512 \\
$d$ & hidden state dimension & 512 \\
$d_{\text{ff}}$ & feed-forward dimension & 2,048 \\
$d_{\bm{z}}$ & latent condition dimension & 128 \\
$d_{\bm{a}}$ & attribute embedding dimension (each) & 64 \\
\hline
\# params & --- & 79.4 mil.\\
\bottomrule
\end{tabular}
\end{table}

\begin{table}[h]
\centering
\caption{The vocabulary used to represent songs in \textit{LPD-17-cleansed} dataset, which is adopted in Section \ref{sec:part1}.}\label{tab:event-token}
\begin{tabular}{l r r}
\toprule
Event type &  Description  & \# tokens \\
\midrule
\textsc{Bar}&  beginning of a new bar & 1 \\ 
\textsc{Sub-beat}&  position in a bar, in 32nd note steps (\musThirtySecond) & 32 \\ 
\textsc{Tempo}&  32$\sim$224 bpm, in steps of 3 bpm & 64 \\ 
\textsc{Pitch}$^*$&  MIDI note numbers (pitch) 0$\sim$127 & 1,757 \\ 
\textsc{Velocity}$^*$&  MIDI velocities 3$\sim$127 & 544 \\ 
\textsc{Duration}$^*$&  multiples (1$\sim$64 times) of \musThirtySecond  & 1,042 \\
\midrule
\textbf{All events} & --- & \textbf{3,440} \\
\bottomrule
\multicolumn{3}{l}{\footnotesize{$^*$: unique for each of the 17 tracks (instruments)}} \\
\end{tabular}
\end{table}

\begin{table}[h]
\centering
\caption{The vocabulary used to represent piano songs in \textit{AILabs.tw-Pop1K7} dataset, on which \textit{MuseMorphose} (see Section \ref{sec:part2}) is trained.}\label{tab:event-token-remi}
\begin{tabular}{l r r}
\toprule
Event type &  Description  & \# tokens \\
\midrule
\textsc{Bar}&  beginning of a new bar & 1 \\ 
\textsc{Sub-beat}&  position in a bar, in 16th note steps (\musSixteenth) & 16 \\ 
\textsc{Tempo}&  32$\sim$224 bpm, in steps of 3 or 6 bpm & 54 \\ 
\textsc{Pitch}&  MIDI note numbers (pitch) 22$\sim$107 & 86 \\ 
\textsc{Velocity}&  MIDI velocities 40$\sim$86 & 24 \\ 
\textsc{Duration}&  multiples (1$\sim$16 times) of \musSixteenth  & 16 \\
\textsc{Chord}&  chord markings (root \& quality)  & 133 \\
\midrule
\textbf{All events} & --- & \textbf{330} \\
\bottomrule
\end{tabular}
\end{table}

\end{document}